\RequirePackage{fix-cm}
\documentclass[smallextended]{svjour3}       
\smartqed  
\usepackage[misc]{ifsym}

\usepackage{amsmath,amssymb,amsfonts}
\usepackage{graphicx}
\usepackage{textcomp}
\usepackage{xcolor}
\usepackage{multirow}
\usepackage[noend]{algpseudocode}
\usepackage{algorithmicx,algorithm,algpseudocode}
\algdef{SE}[DOWHILE]{Do}{doWhile}{\algorithmicdo}[1]{\algorithmicwhile\ #1}%

\usepackage[square,numbers,sort&compress]{natbib}

\usepackage{bm}
\usepackage{mathrsfs}

\usepackage{subfigure}

\usepackage{lineno}
\usepackage{authblk}

\journalname{Knowledge and Information Systems}

\begin{document}

\title{Efficient Trajectory Compression and Range Query Processing\thanks{This work was supported in part by the National Natural Science Foundation of China under grants No.U19A2059, No.61632010, No.61732003, No.61832003 and No.U1811461 and Key Research and Development Projects of the Ministry of Science and Technology under grant No.2019YFB2101902.}}


\author{Hongbo Yin\textsuperscript{*}	\and 	\Letter \ Hong Gao\textsuperscript{*}	\and	Binghao Wang\textsuperscript{*}	\and	Sirui Li\textsuperscript{*}	\and	Jianzhong Li\textsuperscript{*}}

\institute{Hongbo Yin 		\\ 	hongboyin@hit.edu.cn           
\\   \\     \Letter \ Hong Gao, Corresponding author 		\\  honggao@hit.edu.cn
\\   \\		Binghao Wang 	\\ 	wangbinghao@hit.edu.cn
\\   \\		Sirui Li		\\	kuwylsr@hit.edu.cn
\\   \\		Jianzhong Li	\\	lijzh@hit.edu.cn
\\ 	 \\		\textsuperscript{*}School of Computer Science and Technology, Harbin Institute of Technology, Harbin 150001, China
}

\date{Received: date / Accepted: date}

\maketitle


\begin{abstract}
Nowadays, there are ubiquitousness of GPS sensors in various devices collecting, transmitting and storing tremendous trajectory data. However, such an unprecedented scale of GPS data has posed an urgent demand for not only an effective storage mechanism but also an efficient query mechanism. Line simplification in online mode, searving as a mainstream trajectory compression method, plays an important role to attack this issue. But for the existing algorithms, either their time cost is extremely high, or the accuracy loss after the compression is completely unacceptable. To attack this issue, we propose $\epsilon \_$Region based Online trajectory Compression with Error bounded (ROCE for short), which makes the best balance among the accuracy loss, the time cost and the compression rate. The range query serves as a primitive, yet quite essential operation on analyzing trajectories. Each trajectory is usually seen as a sequence of discrete points, and in most previous work, a trajectory is judged to be overlapped with the query region $R$ iff there is at least one point in this trajectory falling in $R$. But this traditional criteria is not suitable when the queried trajectories are compressed, because there may be hundreds of points discarded between each two adjacent points and the points in each compressed trajectory are quite sparse. And many trajectories could be missing in the result set. To address this, in this paper, a new criteria based on the probability and an efficient Range Query processing algorithm on Compressed trajectories RQC are proposed. In addition, an efficient index \emph{ASP\_tree} and lots of novel techniques are also presented to accelerate the processing of trajectory compression and range queries obviously. Extensive experiments have been done on multiple real datasets, and the results demonstrate superior performance of our methods.
\end{abstract}

\section{Introduction}\label{section:Introduction}
With the unprecedented growth of GPS-equipped devices, such as smart-phones, vehicles and wearable smart devices, massive and increasing volumes of trajectories recording the movements of humans, vehicles, or animals, are being generated for location based services, trajectory mining, wildlife tracking or other useful applications. For example, DiDi Chuxing is China's largest online ridesharing platform. It needs to process up to fifty million trip requests in a single day$\footnote{https://tech.sina.com.cn/roll/2020-08-26/doc-iivhvpwy3125825.shtml}$, i.e., up to thousands of requests in a rush second. This also suggests that thousands or even tens of thousands of trajectories are generated per second. However, such an increasing amount of the trajectory data collected brings a great deal of hardship on not only storing but also querying.

As an effective solution to solve the problem, line simplification, a mainstream lossy trajectory compression method, uses a sequence of consecutive line segments with much smaller size to approximately represent the trajectories and has drawn wide attention. The existing line simplification methods fall into two categories, i.e. batch mode and online mode. For each trajectory, algorithms in batch mode, such as Douglas-Peucker\cite{douglas1973algorithms}, SP\cite{long2013direction}, Intersect\cite{long2013direction} and Error-Search\cite{2014Trajectory1}, require that all points in this trajectory must be loaded in the local buffer before compression, which means that the local buffer must be large enough to hold the entire trajectory at least. Thus, the space complexities of these algorithms are at least $O(N)$, or even $O(N^{2})$, where $N$ is the number of input trajectory points. Such high space complexities limit the application of these algorithms in resource-constrained environments, such as the tiny tracking devices on flying foxes, whose RAM barely reaches 4 KBytes\cite{liu2015bounded}. Therefore, more work focuses on the other kind of compression methods, algorithms in online mode, which only need a limited and quite small size of local buffer to compress trajectories in an online processing manner. Thus there are much more application scenarios where algorithms in online mode can be used, such as compressing streaming data. For these algorithms, there is a tradeoff among the execution time, the accuracy loss and the compression rate, which are the three indicators used to measure their performance. And the key issue is how to reach a good balance. As reported in Table \ref{table:CompressionTimeVSPEDError}, part of the experimental results\cite{zhang2018trajectory}, for existing compression algorithms, either the time cost is extremely high, such as BQS and FBQS, or the accuracy loss of the compressed trajectories is totally intolerable, such as Angular, Interval and OPERB. So for algorithms in online mode, it is still a great challenge to compress trajectories with less execution time and less accuracy loss.

\begin{table}
	\centering
	\caption{The time cost and accuracy loss of some compression algorithms in online mode}
	\scalebox{0.8}{
		\begin{tabular}{|c|c|c|c|c|c|}
			\hline
			Compression Algorithm	&  BQS\cite{liu2015bounded,liu2016novel}   								& FBQS\cite{liu2015bounded,liu2016novel}  &  Angular\cite{ke2016online}  &  Interval\cite{ke2017efficient}  &  OPERB\cite{lin2017one}  \\ \hline	
			Execution Time per Point $(\mu s)$ 		&  500.91 & 405.38 &  0.20      &  0.28      &  0.97    \\ \hline
			The Maximum PED Error        					&  38.23  & 36.63  &  1532.65  &  1889.81   &  306.20   \\ \hline
	\end{tabular}}
	\label{table:CompressionTimeVSPEDError}
\end{table}

To attack this issue, we propose a new online line simplification compression algorithm ROCE with only $O(N)$ time complexity and $O(1)$ space complexity, which makes the best balance among the accuracy loss, the time cost and the compression rate. Among the fastest algorithms, the accuracy loss of the compressed trajectories generated by ROCE is always the smallest, and among algorithms with the smallest accuracy loss, ROCE is always the fastest.

\begin{figure}
	\centering
	\includegraphics[width=3.5in]{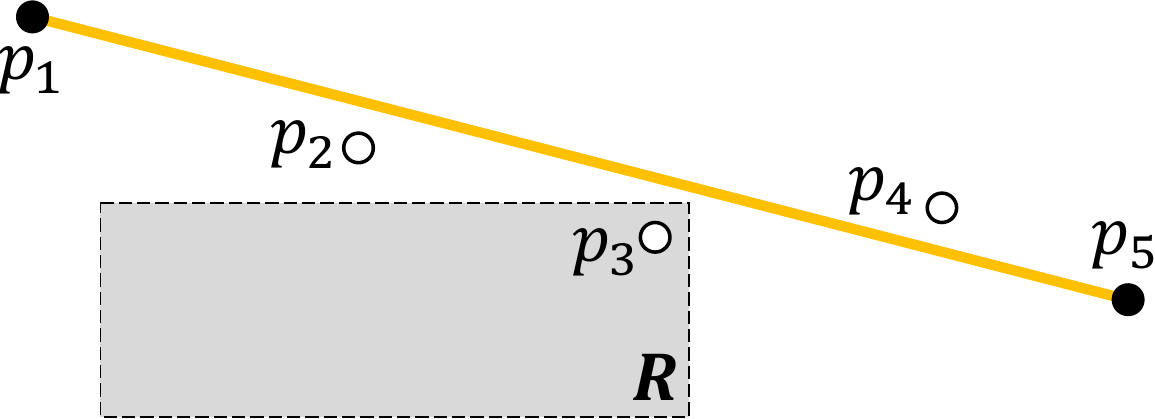}
	\caption{The discarded points $p_{3}$ falls in query region $R$}
	\label{figure:LineSegmentOverlapsR}
\end{figure}

Compressing trajectories can reduce not only the cost of storage and transmission, but also the cost of queries greatly. Large trajectory data facilitates various real-world applications, such as trajectory pattern mining, route planing and travel time prediction. For these various applications, there is a type of trajectory queries named range queries, serving as a primitive, yet essential operation. The previous work related to range queries, such as \cite{zhang2018trajectory,zhang2018gpu,dong2018gat}, usually see each trajectory as a sequence of discrete points, and a trajectory is regarded to be overlapped with the query region $R$ iff there exist one point in this trajectory falls in $R$. However, this traditional criteria is completely unsuitable for range quering on compressed trajectories, and many trajectories will be missing in the result set. Because there may be hundreds of points discarded between each two adjacent points in compressed trajectories and the points in each compressed trajectory are extremely sparse. If some points in a trajectory fall in the query region, but these points are discarded after the compression, such as the situation shown in Figure \ref{figure:LineSegmentOverlapsR}, then in the final result set of the range query, such a trajectory is missing. To solve this problem, we propose a specially designed criteria about range queries on compressed trajectories and an effective algorithm about how to process range queries on compressed trajectories with just a little additional information. And the difference between the range query result on compressed trajectories and that on the corresponding raw trajectories can be reduced greatly.

The main contributions of our work are listed as follows:

\begin{itemize}
	\item Point-to-Segment Euclidean Distance (PSED), a more reasonable accuracy loss metric, is defined to measure the degree of the accuracy loss after a trajectory is compressed. 
	
	\item Based on PSED, we propose a new online line simplification compression algorithm ROCE with only $O(N)$ time complexity and $O(1)$ space complexity, which achieves the best balance among the accuracy loss, the time cost and the compression rate.
	
	\item For range queries on compressed trajectories, a new criteria based on the probability and a new range query processing algorithm RQC is proposed to reduce the difference between the query results on compressed trajectories and on the corresponding raw trajectories greatly. An efficient index \emph{ASP\_tree} is also presented to accelerate the processing of range queries greatly.
	
	\item Extensive comparison experiments were conducted on real-life trajectory datasets, and the results demonstrate the superior performance of our methods.
\end{itemize}

The rest of this paper is organized as follows. We present a new accuracy loss metric PSED and the compression algorithm ROCE in Section \ref{section:ROCECompressionAlgorithm}. The index \emph{ASP\_tree} and the range query processing algorithm RQC are introducted in Section \ref{section:RangeQuery}. Section \ref{section:ExperimentalEvalution} shows the detailed experimental results and the corresponding analysis. Section \ref{section:RelatedWork} reviews the related works, and Section \ref{section:Conclusions} concludes our work.

\section{ROCE Compression Algorithm}\label{section:ROCECompressionAlgorithm}

In this section, a more reasonable accuracy loss metric PSED is proposed first. Then based on PSED, a new compression algorithm ROCE, which makes the best balance among the accuracy loss, the time cost and the compression rate, is introduced in detail.

\subsection{Basic Concepts and Notations}\label{subsection:BasicNotations}

\begin{definition}(Trajectory $T$): A trajectory $T$ can be expressed as a sequence of discrete points $\{p_{1}, p_{2}, ..., p_{N}\}$, where $T[i]=p_{i}(x,y,t)$ means that the moving object was located at longitude $x$ and latitude $y$ at time $t$. And $\forall 1 \leq i \leq j \leq N$, $p_{i}.t<p_{j}.t$.\end{definition}

Given a trajectory $T=\{p_{1}, p_{2}, ..., p_{N}\}$, $\forall i,j(1 \leq i < j \leq N)$, $T[i:j]=\{p_{i}, p_{i+1}, ..., p_{j}\}$ represents a trajectory segment with $(j-i+1)$ consecutive points. And the line segment $p_{i}p_{j}$ can approximately represent such a trajectory segment, i.e., $p_{i}p_{j}$ is the compressed form of $T[i:j]$. $p_{i+1}$. $p_{i+2}$, ..., $p_{j-1}$ are called the discarded points, and $p_{i}p_{j}$ is called the corresponding line segments of $p_{i}$, $p_{i+1}$, ..., $p_{j}$. 

For a trajectory $T$, a compression algorithm is to divide $T$ into a sequence of consecutive trajectory segments $\{T[i_{1}:i_{2}], T[i_{2}:i_{3}], ..., T[i_{n-1}:i_{n}]\}(i_{1}=1, i_{n}=N)$, and each trajectory segment $T[i_{k}:i_{k+1}]$ is approximately represented by the line segment $p_{i_{k}}p_{i_{k+1}}$. Then the corresponding compressed trajectory $T'$ of $T$ consists of a sequence of $n-1$ consecutive line segments $p_{i_{1}}p_{i_{2}}, p_{i_{2}}p_{i_{3}}, ..., p_{i_{n-1}}p_{i_{n}}$, and $T'$ is denoted as $\{p_{i_{1}}, p_{i_{2}},..., p_{i_{n}}\}(i_{1}=1, i_{n}=N)$ to simplify the representation. These consecutive line segments approximately describe the movement of the moving object. In order to distinguish an uncompressed trajectory from its corresponding compressed trajectory, we call it a raw trajectory in the following.

\begin{definition}(Compression Rate): Given a compressed trajectory, $T'=\{p_{i_{1}},\\ p_{i_{2}},..., p_{i_{n}}\}(i_{1}=1, i_{n}=N)$ with $n-1$ consecutive line segments, and its corresponding raw trajectory $T=\{p_{1}, p_{2},..., p_{N}\}$ with $N$ points, the compression rate is defined as: \begin{displaymath}r=\frac{N}{n}.\end{displaymath}\end{definition}

\subsection{Accuracy Loss Metric}\label{subsection:PEDandRPED}

After compression, a set of consecutive line segments is used to approximately represent a raw trajectory. When the compression rate is fixed, for a compression algorithm, the smaller accuracy loss of compressed trajectories, the better. And how to measure the accuracy loss calls for a reasonable enough metric. Usually, the accuracy loss of a compressed trajectory is calculated based on the deviation between each discarded point and its corresponding line segment.

Perpendicular Euclidean Distance (PED for short), an accuracy loss metric adopted by most line simplification methods\cite{lin2017one}, e.g. OPW\cite{keogh2001online}, OPW-TR\cite{meratnia2004spatiotemporal}, BQS\cite{liu2015bounded,liu2016novel} and OPERB\cite{lin2017one}, is formally defined as:

\begin{definition}\label{definition:PED}(PED): Given a trajectory segment $T[s:e](s<e)$ and the line segment $p_{s}p_{e}$, the compressed form of $T[s:e](s<e)$, for any discarded point $p_{m}(s<m<e)$ in $T[s:e]$, the PED of $p_{m}$ is calculated as:
	\begin{displaymath}\label{equation:PED} PED(p_{m})=\frac{||\overrightarrow{p_{s}p_{m}} \times \overrightarrow{p_{s}p_{e}}||}{||\overrightarrow{p_{s}p_{e}}||}\end{displaymath} 
	where $\times$ and $|| \ ||$ are respectively to calculate the results of cross product and the length of a vector.\end{definition}

\begin{figure}
	\centering
	\includegraphics[width=4in]{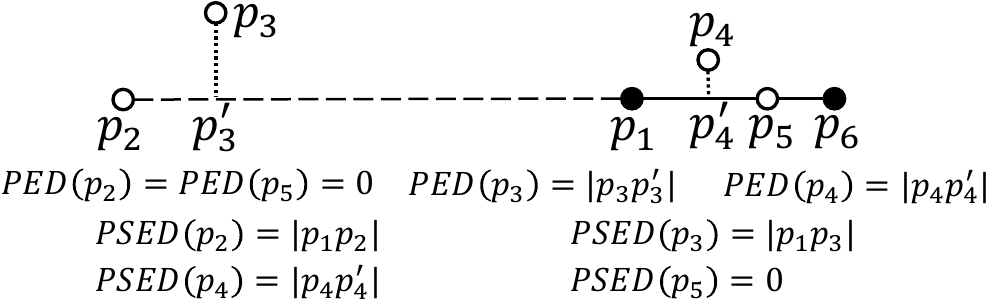}
	\caption{An example shows how to calculate PED and PSED}
	\label{figure:PED&RPED}
\end{figure}

PED measures the deviation between each discarded point and its corresponding line segment by using the shortest Euclidean distance from the discarded point to the straight line on which the corresponding line segment lies. However, PED can hardly describe the deviation accurately when the moving direction changes sharply. For example, it is a quite common situation that active tracked animals or wandering tourists always change their moving direction sharply. Figure \ref{figure:PED&RPED} illustrates the tracked object makes a U-turn, and the line segment $p_{1}p_{6}$ approximately represents the trajectory segment $T[1:6]$ after the compression. The accuracy losses of $p_{2}$ and $p_{3}$ in PED are respectively 0 and $|p_{3}p_{3}'|$. But in fact, $p_{2}$ is obviously far away from the line segment $p_{1}p_{6}$ and the distance between $p_{3}$ and the $p_{1}p_{6}$ is also far more than $|p_{3}p_{3}'|$. The reason for these is that $PED(p_{2})$ and $PED(p_{3})$ are both calculated based on the perpendicular distance between the discarded points and the extension line of $p_{1}p_{6}$. Thus the compressed trajectories, which are generated by the compression algorithms whose accuracy loss metric is PED, are not able to reflect the real movement patterns.

To attack this issue, we define a more reasonable accuracy loss metric PSED to measure the accuracy loss after the compression. The key difference between PSED and PED is that PSED adopts the shortest Euclidean distance from a point to its corresponding line segment, rather than the straight line on which the corresponding line segment lies. PSED is formally defined as follows:

\begin{definition}\label{definition:RPED}(PSED): Given a trajectory segment $T[s:e](s<e)$ and the line segment $p_{s}p_{e}$, the compressed form of $T[s:e](s<e)$, for any discarded point $p_{m}(s<m<e)$ in $T[s:e]$, the PSED of $p_{m}$ is calculated according to the following cases:
	\begin{displaymath}\begin{aligned}PSED(p_{m})=\left\{	
			\begin{tabular}{c l}
				\multirow{2}{*}{\begin{Large}$\frac{||\overrightarrow{p_{s}p_{m}} \times \overrightarrow{p_{s}p_{e}}||}{||\overrightarrow{p_{s}p_{e}}||}$\end{Large},}  & \multirow{2}{*}{$\overrightarrow{p_{s}p_{m}} \cdot \overrightarrow{p_{s}p_{e}} \geq 0 \ {\rm and} \ \overrightarrow{p_{m}p_{e}} \cdot \overrightarrow{p_{s}p_{e}} \geq 0$} \\ \\
				$min(||\overrightarrow{ p_{s}p_{m} }||,||\overrightarrow{ p_{m}p_{e} }||)$, & otherwise  \\
			\end{tabular}
			\right.\end{aligned}\end{displaymath}
	where $\times$ and $\cdot$ are respectively to calculate the results of cross product and dot product.\end{definition}

In Definition \ref{definition:RPED}, that both $\overrightarrow{p_{s}p_{m}} \cdot \overrightarrow{p_{s}p_{e}} \geq 0$ and $\overrightarrow{p_{m}p_{e}} \cdot \overrightarrow{p_{s}p_{e}} \geq 0$ are satisfied means that the perpendicular point of $p_{m}$ falls on the line segment $p_{s}p_{e}$. In Figure \ref{figure:PED&RPED}, since the perpendicular points of $p_{2}$ and $p_{3}$ both fall on the extension line of $p_{1}p_{6}$, $PSED(p_{2})=min(|p_{1}p_{2}|,|p_{2}p_{6}|)=|p_{1}p_{2}|$ and $PSED(p_{3})=min(|p_{1}p_{3}|,|p_{3}p_{6}|)=|p_{1}p_{3}|$. For $p_{4}$ and $p_{5}$, whose corresponding perpendicular points are both on the line segment $p_{1}p_{6}$, $PSED(p_{4})=|p_{4}p_{4}'|$ and $PSED(p_{5})=0$. 

Based on PSED, the $\epsilon$-error-bounded compressed trajectory is defined as follows:

\begin{definition}\label{definition:Error-boundedTrajectory}($\epsilon$-error-bounded Compressed Trajectory): Given a threshold value $\epsilon$, a compressed trajectory $T'=\{p_{i_{1}}, p_{i_{2}},..., p_{i_{n}}\}(p_{i_{1}}=p_{1},p_{i_{n}}=p_{N})$ and its corresponding raw trajectory $T=\{p_{1}, p_{2},..., p_{N}\}$. If $\forall p_{m} \in T$, $PSED(p_{m}) \leq \epsilon$, then $T'$ is $\epsilon$-error-bounded, and $\epsilon$ is an upper bound of PSED.\end{definition}

\subsection{Algorithm ROCE}\label{subsection:AlgorithmROCE}

In this part, a new compression algorithm in online mode named ROCE, which makes the best balance among the accuracy loss, the time cost and the compression rate, is presented. Given a raw trajectory $T=\{p_{1}, p_{2}, ..., p_{N}\}$ and the upper bound of PSED $\epsilon$, by adopting a greedy strategy, ROCE is to compress $T$ into an $\epsilon$-error-bounded compressed trajectory $T'$, which consists of a sequence of consecutive line segments. 

In order to determine whether a compressed trajectory is $\epsilon$-error-bounded or not much more conveniently, we define a new concept $\epsilon \_$Region as below:

\begin{definition}\label{definition:epsilon-Region}($\epsilon \_$Region $C_{i}$): Given a raw trajectory point $p_{i}$ and the upper bound of PSED $\epsilon$, the circle region $C_{i}$, whose center and radius are respectively $p_{i}$ and $\epsilon$, is called the $\epsilon \_$Region of $p_{i}$.\end{definition}

Then, it is quite easy to get the following property about $\epsilon \_$Region:

\begin{lemma}\label{lemma:PointAndCircle1}
	A trajectory segment $T[s:e](s<e)$ is compressed into a line segment $p_{s}p_{e}$. For any discarded point $p_{m}(s<m<e)$ in $T[s:e]$, $PSED(p_{m}) \leq \epsilon$, where $\epsilon$ is the upper bound of PSED, iff $p_{s}p_{e}$ intersects $C_{m}$. $p_{s}p_{e}$ is $\epsilon$-error-bounded iff $p_{s}p_{e}$ intersects all $\epsilon \_$Regions of discarded points, i.e. $C_{s+1}$, $C_{s+2}$, ..., $C_{e-1}$.\end{lemma}

\begin{figure}
	\centering
	\includegraphics[width=4.5in]{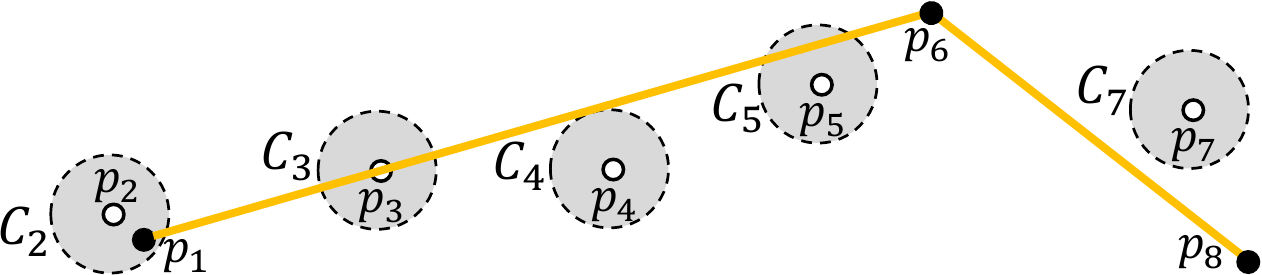}
	\caption{$T'=\{p_{1}, p_{6}, p_{8}\}$ is a compressed trajectory of $T=\{p_{1},p_{2},...,p_{8}\}$}
	\label{figure:TheStartingPonitOfROCE}
\end{figure}

As shown in Figure \ref{figure:TheStartingPonitOfROCE}, the raw trajectory $T=\{p_{1},p_{2},...,p_{8}\}$ is compressed into $T'$, which consists of two line segments, i.e. $p_{1}p_{6}$ and $p_{6}p_{8}$. For any discarded point in the trajectory segment $T[1:6]$, $p_{1}p_{6}$ intersects its corresponding $\epsilon \_$Region. Thus, $p_{1}p_{6}$ is clearly $\epsilon$-error-bounded. It is obvious that the line segment $p_{6}p_{8}$ does not intersect the corresponding $\epsilon \_$Region of $p_{7}$, i.e. $C_{7}$, and $PSED(p_{7})>\epsilon$. Thus neither $p_{6}p_{8}$ nor $T'$ is $\epsilon$-error-bounded.

Given the upper bound of PSED $\epsilon$ and a raw trajectory $T=\{p_{1}, p_{2},..., p_{N}\}$, an optimal compression is to compress $T$ into an $\epsilon$-error-bounded trajectory $T'$ consisting of the smallest number of consecutive line segments. $T$ can be divided into $2^{N-1}$ different sets of consecutive trajectory segments, which means that there are up to $2^{N-1}$ different compressed strategies and the search space is exponential. By adopting a greedy strategy and some effective tricks, ROCE, an efficient approximate algorithm, compresses trajectories in an online processing manner. The first thing, ROCE anchors the start point $p_{s}$ of a trajectory segment to be compressed. $p_{f}$, where $f$ is a variable and initialized to $(s+2)$, is selected as the current float point. Then by using $p_{s}$ and $p_{f}$ a trajectory segment $T[s:f]$ is defined. $p_{f+1}$ is assigned as the new float point and $(f+1)$ is assigned as $f$, if $\forall p_{m}(s<m<f) \in T[s:f]$, $PSED(p_{m}) \leq \epsilon$. Otherwise, $T[s:f-1]$ is compressed into a line segment $p_{s}p_{f-1}$, and the anchor point of the next trajectory segment to be compressed is set to $p_{f-1}$.

Each time ROCE checks whether the last float point $p_{f-1}$ is the final end point of the current trajectory segment or not, each point $p_{m}(s<m<f)$ needs to be scanned to calculate the corresponding PSED to verify whether the line segment $p_{s}p_{f}$ is $\epsilon$-error-bounded. So each point needs to be scanned many times during the compression, and lots of execution time is wasted here. To attack this issue, the candidate region is adopted by ROCE, and each point just needs to be scanned only once. $(p_{s},p_{f})\_Region$ and $T[s:f]\_CandidateRegion$ are formally defined as follows:

\begin{definition}($(p_{s},p_{f})\_Region$): Given a trajectory segment $T[s:f](s<f$ and  $|p_{s}p_{f}|>\epsilon)$ and the upper bound of PSED $\epsilon$. Then $p_{s}$ is outside the $\epsilon \_Region$ $C_{f}$ of $p_{f}$, and two tangent rays of $C_{f}$ starting from $p_{s}$ named $tr_{s,f}$ and $tr_{s,f}'$ can be gotten. The minor sector enclosed by $tr_{s,f}$ and $tr_{s,f}'$, excluding the circular region, whose center and radius are $p_{s}$ and $|p_{s}p_{f}|$ respectively, is called $(p_{s},p_{f})\_Region$.\end{definition}

\begin{definition}($T[s:f]\_CandidateRegion$): Given a trajectory segment $T[s:f](s<f$ and $|p_{s}p_{f}|>\epsilon)$ and the upper bound of PSED $\epsilon$. $T[s:f]\_CandidateRegion \\ =  (p_{s},p_{s+1})\_Region \bigcap$ $(p_{s},p_{s+2})\_Region \bigcap$ ... $\bigcap$ $(p_{s},p_{f})\_Region$, i.e., $T[s:f]\_CandidateRegion = T[s:f-1]\_CandidateRegion \bigcap$ $(p_{s},p_{f})\_Region$ if $s<(f-1)$.\end{definition}

\begin{figure}
	\centering
	\includegraphics[width=4.5in]{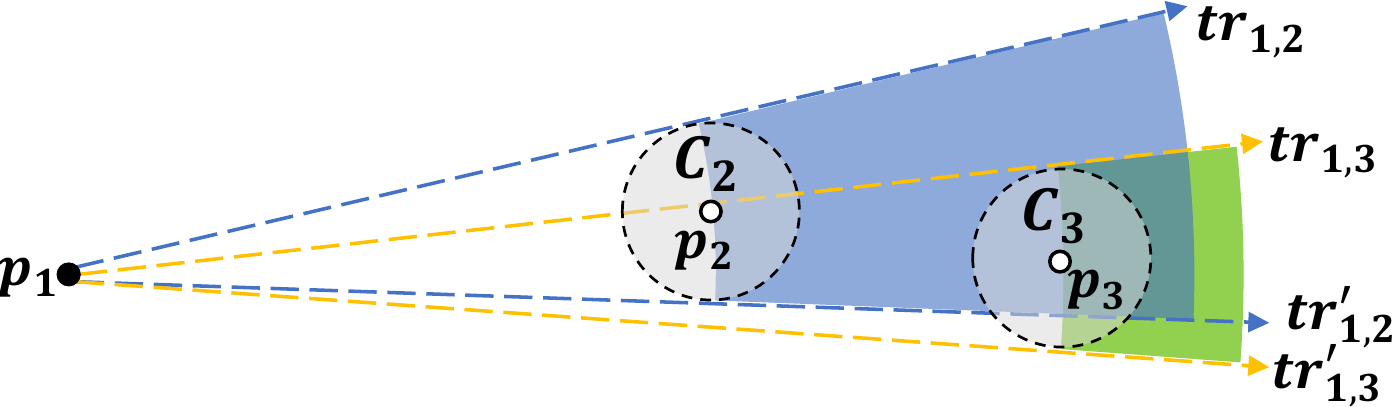}
	\caption{An example shows how to update the candidate region}
	\label{figure:CandidateRegion}
\end{figure}

During the procedure of finding which is the final end point of the current trajectory segment starting from $p_{s}$ to be compressed, if the float point $p_{f+1}$ falls in $T[s:f]\_CandidateRegion$, then $\forall p_{m}(s<m<f+1)$, $PSED(p_{m}) \leq \epsilon$. So by using the candidate region in ROCE, PSED no longer needs to be calculated any more, and each point just needs to be scanned only once to update the current candidate region. Figure \ref{figure:CandidateRegion} gives us an example to show how to update the candidate region. Since $p_{1}$ is outside the $\epsilon \_$Region $C_{2}$ of $p_{2}$, we can get two tangent rays $tr_{1,2}$ and $tr_{1,2}'$ of $C_{2}$. Both $(p_{1},p_{2})\_Region$ and $T[1:2]\_CandidateRegion$ are the region in blue. $p_{3}$ falls in $T[1:2]\_CandidateRegion$. Similarly, $(p_{1},p_{3})\_Region$ is the region in green. Then $T[1:3]\_CandidateRegion$ is the overlapping region of $T[1:2]\_CandidateRegion$ and $(p_{1},p_{3})\_Region$. According to Lemma \ref{lemma:PointAndCircle1}, the line segment $p_{1}p_{4}$ is $\epsilon$-error-bounded iff the next point $p_{4}$ falls in $T[1:3]\_CandidateRegion$, because the line segment $p_{1}p_{4}$ must intersect all $\epsilon \_$Regions of discarded points, i.e., $C_{2}$ and $C_{3}$.

The pseudo code of ROCE is formally introduced in Algorithm \ref{algorithm:ROCE}. Starting from the first point, points in the trajectory are scanned one by one. In each iteration, ROCE tries to find which is the final end point of the current trajectory segment to be compressed, and this trajectory segment is compressed into a line segment by ROCE (Line \ref{$StartPoint=T[i]$}-\ref{algorithmLable8}). For the following points of the start point, if none of their distances to the start point are more than $\epsilon$, for any line segment starting from the start point, it must intersect all their corresponding $\epsilon \_$Regions. Thus according to Lemma \ref{lemma:PointAndCircle1}, their restrictions no longer need to be thought about (Line \ref{algorithmLable2}-\ref{algorithmLable3}).

\begin{algorithm}[ht]
	\caption{: ROCE Algorithm}
	\label{algorithm:ROCE} 
	\hspace*{0.02in} {\bf Input:} 
	a raw trajectory $T=\{p_{1}, p_{2},..., p_{N}\}$ and the upper bound of PSED $\epsilon$\\
	\hspace*{0.02in} {\bf Output:} 
	an $\epsilon$-error-bounded compressed trajectory $T'$ of $T$
	\begin{algorithmic}[1]		
		\State {$i \leftarrow 1$, $T' \leftarrow [T[1]]$}
		
		\While {$i \leq N$}
		\State {$StartPoint \leftarrow T[i]$}\label{$StartPoint=T[i]$}	
		\State {$CandidateRegion.initialize(StartPoint)$}
		\State {$i \leftarrow i+1$}
		
		\While {$(dist(StartPoint,T[i]) \leq \epsilon)$ and $(i \leq N$)}\label{algorithmLable2}
		\State {$i \leftarrow i+1$}
		\EndWhile\label{algorithmLable3}
		
		\While {($T[i]$ in $CandidateRegion$) and ($i \leq N$)}\label{algorithmLable4}
		\State {$CandidateRegion.update(T[i],\epsilon)$}
		\State {$i \leftarrow i+1$}
		\EndWhile\label{algorithmLable5}
		\State {$i \leftarrow i-1$}\label{algorithmLable6}
		\State {$T'.append(T[i])$}\label{algorithmLable7}\label{algorithmLable8}
		\EndWhile
		\Return $T'$
	\end{algorithmic}
\end{algorithm}

By using the candidate region, ROCE is a one-pass error bounded trajectory compression algorithm, since each point just needs to be scanned onle once to update the current candidate region. So the time complexity of ROCE is $O(N)$. The space complexity of ROCE is only $O(1)$, since only constant and small space is needed by ROCE, no matter how many points to be compressed into a line segment.

\section{Range Query Processing}\label{section:RangeQuery}

In most previous work\cite{zhang2018trajectory,zhang2018gpu,dong2018gat}, each raw trajectory is usually regarded as a sequence of discrete points, and a raw trajectory is determined to be overlapped with the query region $R$ iff at least one point in this trajectory falls in $R$. But this traditional criteria is not suitable for range queries on compressed trajectories, since it will lead to many trajectories are missing in the result set as discussed in Section \ref{section:Introduction}. To address this, we propose a new criteria based on the probability about range queries on compressed trajectories and an effective algorithm about how to process range queries on compressed trajectories. The new criteria is formally defined as:

\begin{definition}\label{definition:RangeQuery}(Range Query on Compressed Trajectories): Given a query region $R$, a compressed trajectory dataset $\mathbb{T'}$ and a probability threshold value $p$, the range query result $Q_{r}(R,\mathbb{T'},p)$ consists of all such compressed trajectories in $\mathbb{T'}$, the probabilities P of whose corresponding raw trajectories are overlapped with $R$ are all larger than $p$, i.e. \begin{displaymath}\begin{aligned}Q_{r}(R,\mathbb{T'},p)= \{ & T' \in \mathbb{T'}| P(\exists p_{i} \in RawTrajectory(T'),\ s.t. &p_{i} \in R)>p\}\end{aligned}\end{displaymath}\end{definition}

\begin{figure}
	\centering
	\includegraphics[width=4.2in]{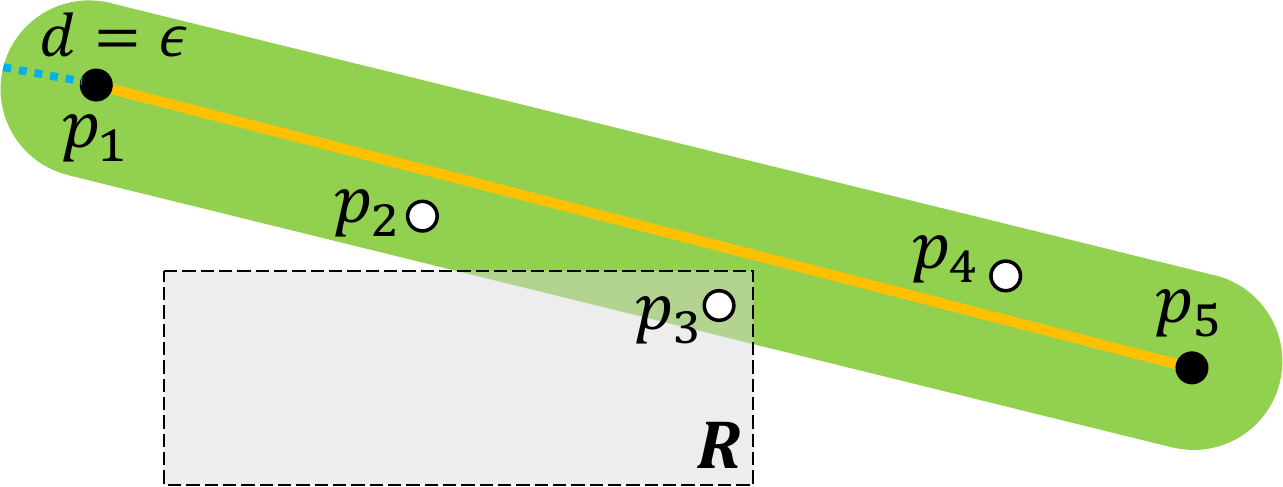}
	\caption{The trajectory segment $T[1:5]$ is compressed into a line segment $p_{1}p_{5}$}
	\label{figure:E-RegionOverlapsR}
\end{figure}

Though query regions are considered as two-dimensional rectangles for simplicity, our method can be easily adapted to handling query regions in arbitrary shapes. As shown in Figure \ref{figure:E-RegionOverlapsR}, the trajectory segment $T[1:5]$ is compressed into a line segment $p_{1}p_{5}$ by ROCE algorithm with the upper bound of PSED $\epsilon$. The only certainty is that there are 3 points discarded between $p_{1}$ and $p_{5}$, and these 3 discarded points are all within the green region, named $\epsilon\_Bounding\ Region$ ($\epsilon\_BR$ for short), which is formally defined as:

\begin{definition}\label{definition:epsilon_BR}($\epsilon\_BR(p_{i}p_{j})$): Given a line segment $p_{i}p_{j}$ and the upper bound of PSED $\epsilon$, $\epsilon\_BR(p_{i}p_{j})$ is the region consists of all points, whose PSEDs to the line segment $p_{i}p_{j}$ are all less than or equal to $\epsilon$.\end{definition}

For a compressed trajectories, based on the positional relationship between the $\epsilon\_BR$ of each line segment and the query region $R$, the probability of the corresponding raw trajectories overlapped with $R$ can be calculated, which will be introduced more detailedly in Section \ref{subsection:RangeQuerybasedonLineSegments}.

However, there are multiple consecutive line segments in each compressed trajectory, and for a range query, it will cost too much to judge the relationship between the query region $R$ and the $\epsilon\_BR$ of each line segment in all compressed trajectories. To address this, we find that for a line segment after compression, if its corresponding $\epsilon\_BR$ and the query region $R$ are not overlapped, then any discarded point approximately represented by this line segment must not fall in $R$. And based on this, RQC follows a filtering-and-verification framework. The filtering step can prune most invalid trajectories at quite low computation cost. In Section \ref{subsection:SpatialIndexASP-tree}, Adaptive Spatial Partition quadtree like index (\emph{ASP\_tree} for short), a high efficient index, is proposed to accelerate the filtering step greatly. Then in Section \ref{subsection:RangeQuerybasedonLineSegments}, the processing procedure of RQC is described in detail.

\subsection{Trajectory Index ASP\_tree}\label{subsection:SpatialIndexASP-tree}
The root node of \emph{ASP\_tree} represents all compressed trajectories falling in the entire region. If there are more than $\xi$ endpoints of all line segments falling in the corresponding region of each node in \emph{ASP\_tree}, where $\xi$ is a threshold value estimated through experiments, then this node is a non-leaf node with 4 child nodes. Otherwise, this node serves as a leaf node. So $\xi$ controls the height of \emph{ASP\_tree}.

To reduce the space overhead of \emph{ASP\_tree}, the detailed information of compressed trajectories is only stored in leaf nodes in the form of $ChildRegion\_ChildPointer$, where $ChildRegion\_ChildPointer$ refers to the corresponding regions and addresses of its 4 child nodes. Each leaf node in \emph{ASP\_tree} stores information in the form of $ID\_LineSegments$. $ID\_LineSegments$ refers to some consecutive line segments of a compressed trajectory whose identifier is $ID$, and the corresponding $\epsilon\_BR$s of these line segments are all overlapped with the corresponding region of this leaf node. So for a compressed trajectory, it may be split into multiple sets of consecutive line segments and stored in different leaf nodes.

\begin{figure}
	\centering
	\includegraphics[width=3.8in]{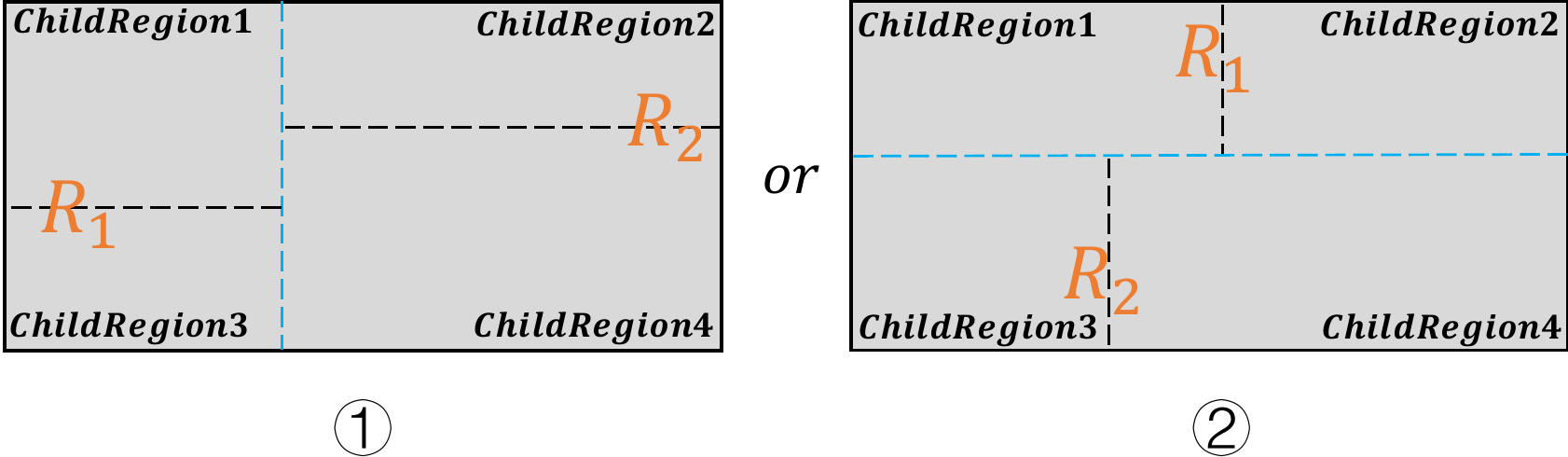}
	\caption{An example shows how to divide the corresponding region of a father node among its 4 child nodes}
	\label{figure:TrajectoryOfNodeIsSplited}
\end{figure}

In a traditional quadtree, if a node is a father node with 4 child nodes, then the corresponding region represented by the father node is evenly divided into four disjoint regions, which are respectively assigned to these 4 child nodes. But this may make the index inclined greatly, which affects the efficiency of the range query processing, because trajectories are not evenly distributed. Thus, it is not suitable to do so. To attack this issue, a data adaptive strategy is adopted in \emph{ASP\_tree}. As shown in Figure \ref{figure:TrajectoryOfNodeIsSplited}, there are totally two ways to divide the corresponding region of a father node. For line segments whose corresponding $\epsilon\_BR$s are overlapped with this region, we first get all endpoints of these line segments falling in this region, and then get the median of all their $x$ dimensions ($y$ dimensions). The median is used to draw a vertical (horizontal) line, which divides this region into two regions named $R_{1}$ and $R_{2}$. After that, the medians of all $y$ dimensions ($x$ dimensions) of all endpoints falling in $R_{1}$ and $R_{2}$ are respectively used to further divide these two regions into four smaller disjoint regions. For these two ways, the way with fewer repeated line segments whose corresponding $\epsilon\_BR$s are overlapped with the four smaller regions will be chosen. The purpose of doing these is to make \emph{ASP\_tree} balanced, which is verified by the experimental results on real-life compressed trajectories in Section \ref{section:ExperimentalEvalution}.

When a region is to be divided into 2 disjoint regions, there may be a special case. For all line segments whose corresponding $\epsilon\_BR$s are overlapped with this region, none of their endpoints fall in this region. In such a case, this region will be divided evenly into 2 smaller regions.

\subsection{Range Query Processing Algorithm}\label{subsection:RangeQuerybasedonLineSegments}

To answer a range query on compressed trajectories, the essential question is how to calculate the probablity of that at least one point in the corresponding raw trajectory of a compressed trajectory falls in the given query region $R$. For each line segment of compressed trajectories, the additional information we can get is how many points are discarded between the two endpoints and that these discarded points are all within the corresponding $\epsilon\_BR$ of this line segment. First of all, we should know what is the probability of that there exists at least one discarded point of its corresponding line segment falls in $R$.

\begin{figure}
	\centering
	\includegraphics[width=3.8in]{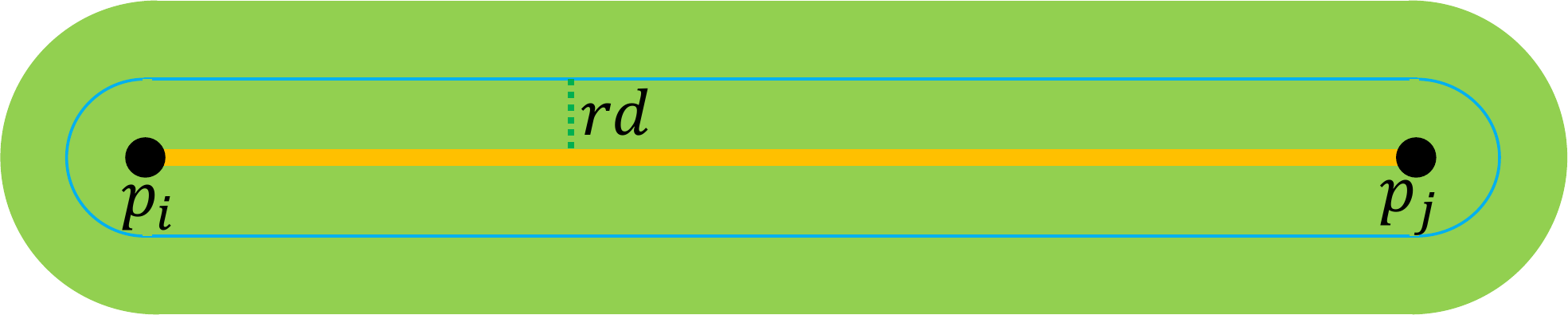}
	\caption{The trajectory segment $T[i:j]$ is compressed into a line segment $p_{i}p_{j}$}
	\label{figure:PointsInE-Region}
\end{figure}

\begin{algorithm}[ht]
	\caption{: Calculating the probablity}
	\label{algorithm:TheProbabilityDiscardedPointsInR}
	\hspace*{0.02in} {\bf Input:} 
	a line segment $p_{i}p_{j}$, the query region $R$, the upper bound of PSED $\epsilon$,	the standard variance $\sigma$ of the Gaussian distribution, the number of sampling points $n_{s}$, the discarded point number $n_{d}$ between $p_{i}$ and $p_{j}$\\
	\hspace*{0.02in} {\bf Output:} 
	the probablity of that at least one discarded point of its corresponding line segment $p_{i}p_{j}$ falls in $R$
	\begin{algorithmic}[1]
		\State {$PointsNumInR \leftarrow 0$}
		\For{$loop \leftarrow 1 ;loop \leq n_{s};loop \leftarrow loop+1$}
		\Do\label{algorithmPDPIR:lable1}
		\State {$rd \leftarrow |generateGaussianDistribution(0,\sigma)|$}
		\doWhile{$rd > \epsilon$}\label{algorithmPDPIR:lable2}
		\State {$SinglePoint \leftarrow samplingOnCurve(p_{i},p_{j},rd)$}\label{algorithmPDPIR:lable3}
		
		\If {$SinglePoint.isInQueryRegion(R)$}
		\State {$PointsNumInR \leftarrow PointsNumInR + 1$}	
		\EndIf
		\EndFor
		
		\State {$rate \leftarrow PointsNumInR/n_{s}$}
		
		\Return $1- (1-rate)^{n_{d}}$\label{algorithmPDPIR:lable4}
	\end{algorithmic}
\end{algorithm}

To address this, we propose Algorithm \ref{algorithm:TheProbabilityDiscardedPointsInR}. After our analysis on lots of real-life compressed trajectories, it is quite clear that for a line segment of a compressed trajectory, its corresponding discarded points are not uniformly distributed in their corresponding $\epsilon\_BR$, but gather near this line segment. For example, for the compressed trajectories whose compression rate is 200, more than 90\% of the PSEDs of discarded points are less than 0.5$\epsilon$ with $\epsilon$ being the upper bound of PSED. So for all compressed trajectories, we choose to use a Gaussian distribution to simulate the distributions of discarded points in all $\epsilon\_BR$s. The mean of the Gaussian distribution is set 0, and the standard variance $\sigma$ can be got and saved when trajectories are being compressed. As shown in Figure \ref{figure:PointsInE-Region}, $rd$ is the absolute value of the PSED randomly generated by using the Gaussian distribution, and we should make sure that $rd \leq \epsilon$ (Line \ref{algorithmPDPIR:lable1}-\ref{algorithmPDPIR:lable2}). For any point on the curve in blue, its PSED to the line segment $p_{i}p_{j}$ is $rd$. On this curve, a point is randomly selected as a representative of discarded points (Line \ref{algorithmPDPIR:lable3}). After $n_{s}$ points are sampled, the probability of that there exists at least one discarded point of its corresponding line segment falls in $R$ can be estimated.

Then based on Algorithm \ref{algorithm:TheProbabilityDiscardedPointsInR}, Algorithm \ref{algorithm:RQC} shows the pseudo code of Range Query processing algorithm on Compressed trajectories RQC. To get the range query result set $Q_{r}(R,\mathbb{T'},p)$, RQC is performed in four steps:

\begin{algorithm}[ht]
	\caption{: RQC Algorithm}
	\label{algorithm:RQC}
	\hspace*{0.02in} {\bf Input:} 
	the query region $R$, the compressed trajectory dataset $\mathbb{T'}$, \emph{ASP\_tree} of $\mathbb{T'}$ and the probability threshold value $p$\\
	\hspace*{0.02in} {\bf Output:} 
	the range query result set $Q_{r}(R,\mathbb{T'},p)$
	\begin{algorithmic}[1]	
		\State {$PQ \leftarrow createPriorityQueue()$}\label{algorithmRQC:lable1}
		\State {$PQ.push($\emph{ASP\_tree}$.root.FourChildNodes())$}
		
		\While {$PQ \neq \phi$}
		\State {$Node \leftarrow PQ.pop()$}
		
		\If {$Node.isOverlap(R)\ and\ !(Node.isLeaf())$}
		\State {$PQ.push(Node.FourChildNodes())$}	
		\EndIf
		
		\If {$Node.isOverlap(R)\ and\ Node.isLeaf()$}
		\State {$S_{c}.append(Node.allID\_LineSegments())$}	
		\EndIf
		\EndWhile\label{algorithmRQC:lable2}

		\State {$(S_{f1},S_{c1}) \leftarrow S_{c}.PruningBasedOnMBR()$}\label{algorithmRQC:lable3}
		
		\State {$(S_{f2},S_{c2}) \leftarrow S_{c1}.verifyByEndpoints()$}\label{algorithmRQC:lable5}
		
		\State {$S_{f3} \leftarrow S_{c2}.verifyBy\epsilon\_BRs(p)$}\label{algorithmRQC:lable4}
		
		\Return $(merge(S_{f1},S_{f2},S_{f3})).getTrajectories(\mathbb{T'})$
	\end{algorithmic}
\end{algorithm}

First (Line \ref{algorithmRQC:lable1}-\ref{algorithmRQC:lable2}), traverse the index \emph{ASP\_tree}, and only all $ID\_LineSegments$ stored in leaf nodes, whose corresponding regions are overlapped with the query region, are left. So by doing this, RQC prunes most invalid compressed trajectories to form a candidate set $S_{c}$, which consists of multiple sequences of consecutive line segments. 

Second (Line \ref{algorithmRQC:lable3}), some efficient pruning strategies based on the $MBR$ (short for the Minimal Bounding Rectangle) are utilized to further reduce the size of the candidate set $S_{c}$, and $S_{c1}$, a much smaller candidate set, and $S_{f1}$, a part of the final result set consisting of multiple $ID$s of compressed trajectories, can be gotten. For a sequence of consecutive line segments in a compressed trajectory, its corresponding $MBR$ is the smallest rectangle which contains all $\epsilon\_BR$s of these line segments. It is clear that if the $MBR$ and the query region $R$ do not overlap, the corresponding points in the corresponding raw trajectory must not fall in $R$. And if the $MBR$ is completely contained in $R$, then there must be at least a point in the corresponding raw trajectory falling in $R$, and this corresponding compressed trajectory must be in the final result set of this range query. By using these two properties, most relationships between sequences of consecutive line segments and the query region $R$ can be determined one by one.

Third (Line \ref{algorithmRQC:lable5}), for each sequence of consecutive line segments in the candidate set $S_{c1}$, whether there exist an endpoint falling in the query region $R$ is determined one by one. And if the answer is yes, then the corresponding raw trajectory must be overlapped with $R$ and the corresponding compressed trajectories must be in the range query result set $Q_{r}(R,\mathbb{T'},p)$. Then the final candidate set $S_{c2}$ and a part of the final result set $S_{f2}$ can be gotten.

Last (Line \ref{algorithmRQC:lable4}), for $S_{c2}$, the final candidate set, Algorithm \ref{algorithm:TheProbabilityDiscardedPointsInR} is used to measure the probablity of that at least one corresponding discarded point of each line segment falls in the query region $R$. For a compressed trajectories $T' = \{p_{i_{1}}, p_{i_{2}},..., p_{i_{n}}\}(i_{1}=1, i_{n}=N)$, supposing that the probablities of its line segments are $r_{1}$, $r_{2}$, ..., $r_{n-1}$ respectively, then $P(T')$, the probablity of the corresponding raw trajectories overlapped with the query region $R$, can be calculated as:
\begin{displaymath}P(T') = 1- \prod \limits_{i=1}^{n-1}(1-r_{i})\end{displaymath}
And if $P(T') > p$ where $p$ is the given probability threshold value, then the $ID$ of $T'$ is put in the final result set $S_{f3}$

\section{Experimental Evalution}\label{section:ExperimentalEvalution}
In this section, the performances of our compression algorithm ROCE, and range query processing algorithm RQC on compressed trajectories are evaluate in detail.

\subsection{Experiment Setup}\label{subsection:ExperimentSetup}

\subsubsection{Datasets}\label{subsubsection:Datasets}
The experiments were conducted on three real-life datasets. The dataset named Animal$\footnote{http://dx.doi.org/10.5441/001/1.78152p3q}$\cite{001/1_78152p3q} records the migrations of 8 young white storks originating from 8 different populations. Because of the tiny tracking devices on these young whites, the sampling rates of these trajectories are relatively low. The dataset named Indoor$\footnote{https://irc.atr.jp/crest2010\_HRI/ATC\_dataset/}$\cite{brvsvcic2013person} records the trajectories of visitors in a shopping center, and the points were sampled very frequently. With a quite large size, the dataset named Planet$\footnote{https://wiki.openstreetmap.org/wiki/Planet.gpx}$ consists of lots of trajectories distributed all over the globe. The movement modes of these trajectories are also pretty rich. These trajectories are sparsely distributed on the Earth, but mainly gather in a large rectangular region, which is about $3.9*10^{5}km^{2}$ in area. All trajectories completely contained in such a large region were selected as a raw dataset called Planet I. The trajectories with less than 1000 points were all removed, since this dataset was to be compressed. In Planet I, there are 96279 raw trajectories and 0.3 billion points in total. The experiments in Section \ref{subsection:RangeQueryProcessingAlgorithmRQC} were all performed on Planet I and its corresponding compressed datasets compressed by ROCE with different compression rates.

Some baseline algorithms are found to be too time-consuming to run on the entire datasets when comparing the execution time of different compression algorithms. So we had to randomly sampled some long trajectories from Animal, Indoor and Planet I. Then we got 3 subsets called Animal II, Indoor II and Planet II with 120, 90 and 47 long trajectories respectively. In these three subsets, there are all about 2 million points in total.

\subsubsection{Experimental Environment}\label{subsubsection:ExperimentalEnvironment}
All experiments were conducted on a linux machine with 32GB memory and a 64-bit, 8-core, 3.6GHz Intel(R) Core (TM) i9-9900K CPU. All algorithms were implemented in C++ on Ubuntu 18.04. Each experiment was repeated over 3 times, and the average is reported below.

\subsection{Performance Evaluation for Compression Algorithms}\label{subsection:TrajectoryCompressionAlgorithms}

Our compression algorithm ROCE was compared with 4 existing compression algorithms in online mode using PED as their error metric, i.e. OPW(BOPW)\cite{keogh2001online,meratnia2004spatiotemporal}, BQS\cite{liu2015bounded,liu2016novel}, FBQS\cite{liu2015bounded,liu2016novel} and OPERB\cite{lin2017one}. For DOTS\cite{cao2017dots}, though its error metric is LISSED but not PED, it was still compared with ROCE, because it was demonstrated to have stable superiority against other compression algorithms in online mode on some indicators\cite{zhang2018trajectory}. The performances of these compression algorithms were measured by the execution time of compression and the accuracy loss of the generated compressed trajectories.

\begin{figure}[ht]
	\centering
	\begin{minipage}[t]{0.495\linewidth}
		\centering
		\subfigure[Animal II]{
			\includegraphics[width=2.25in]{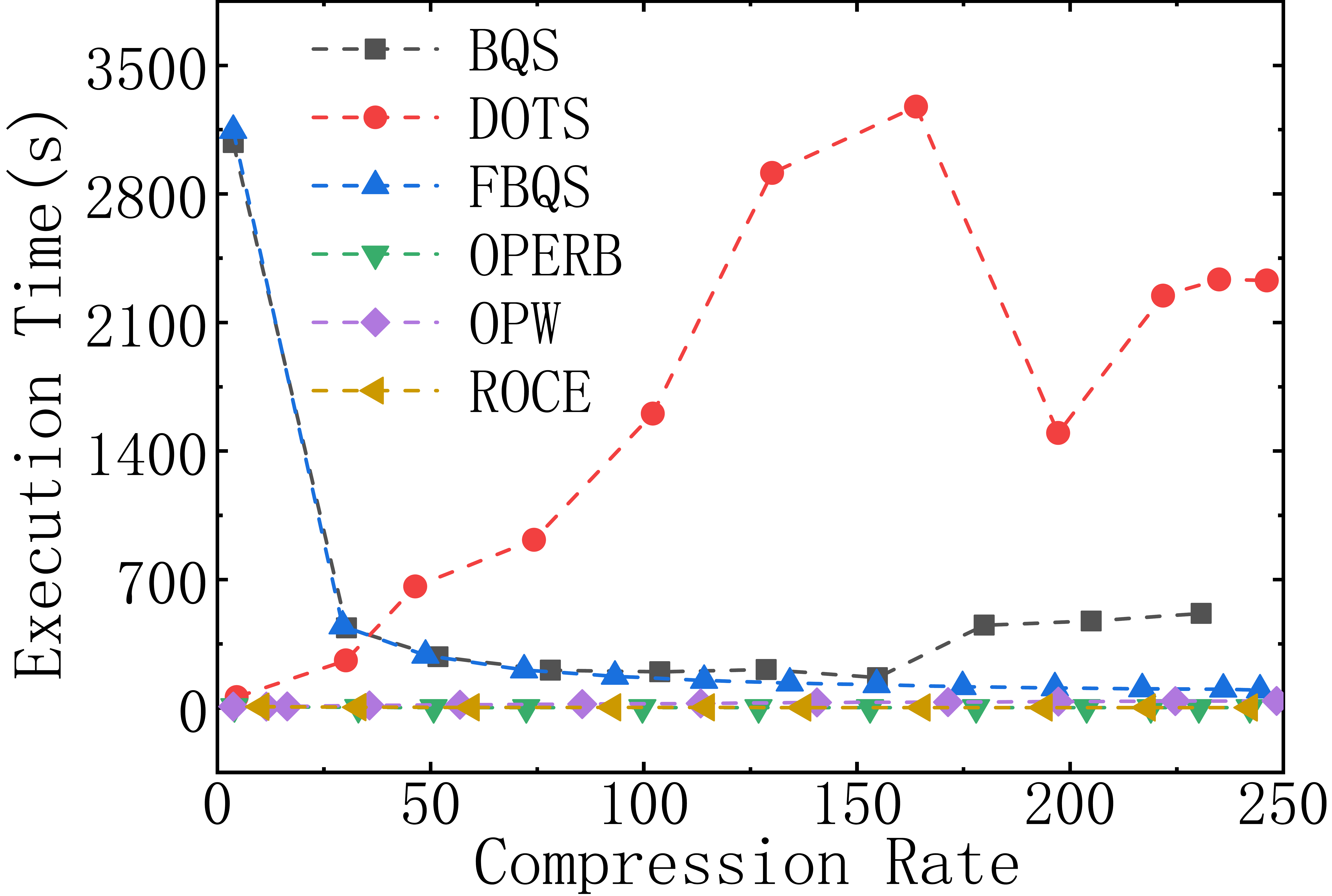}  
		}
	\end{minipage}
	\begin{minipage}[t]{0.495\linewidth}
		\centering
		\subfigure[Indoor II]{
			\includegraphics[width=2.25in]{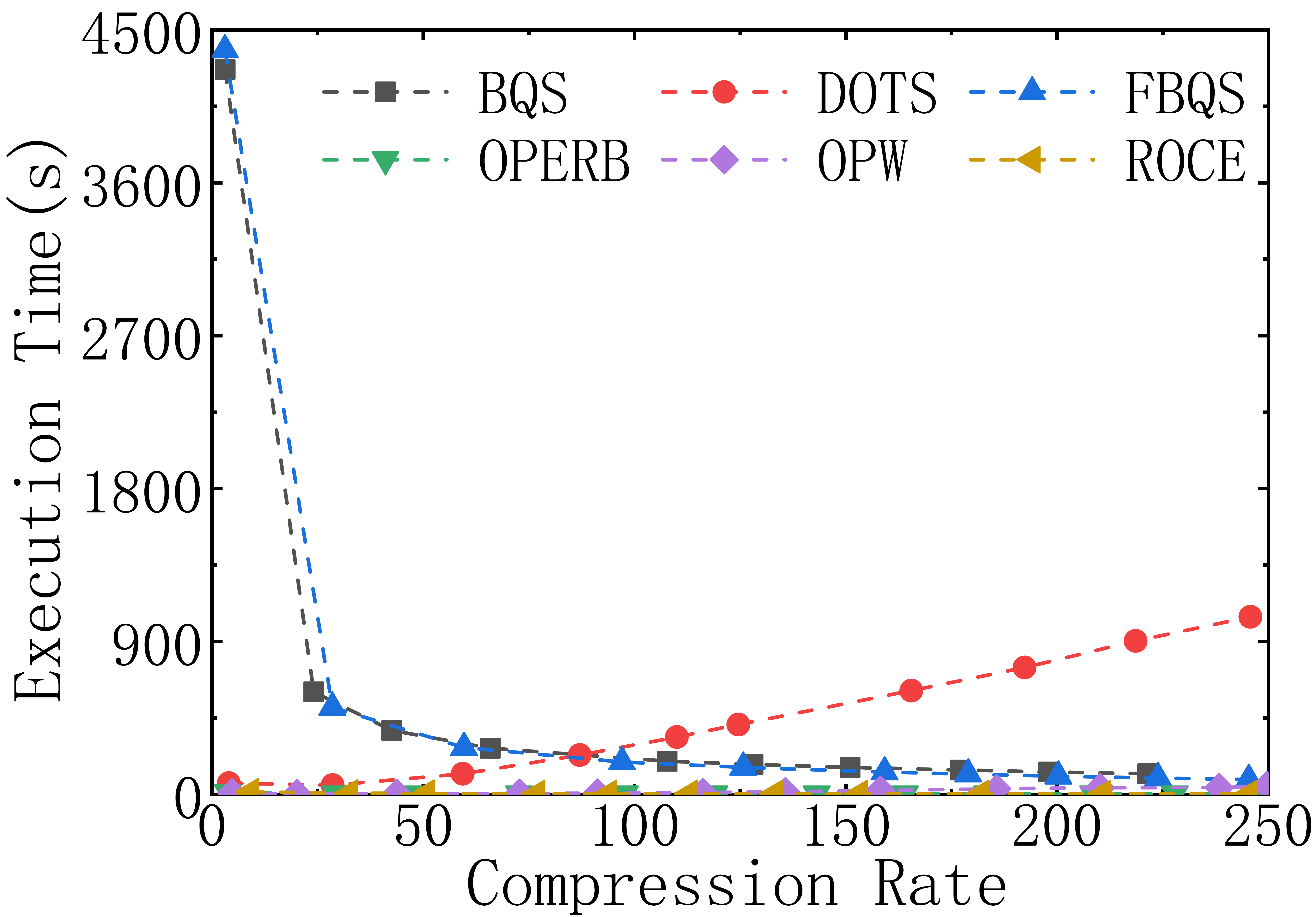}
		}
	\end{minipage}
	\begin{minipage}[t]{0.495\linewidth}
		\centering   
		\subfigure[Planet II]{ 
			\includegraphics[width=2.25in]{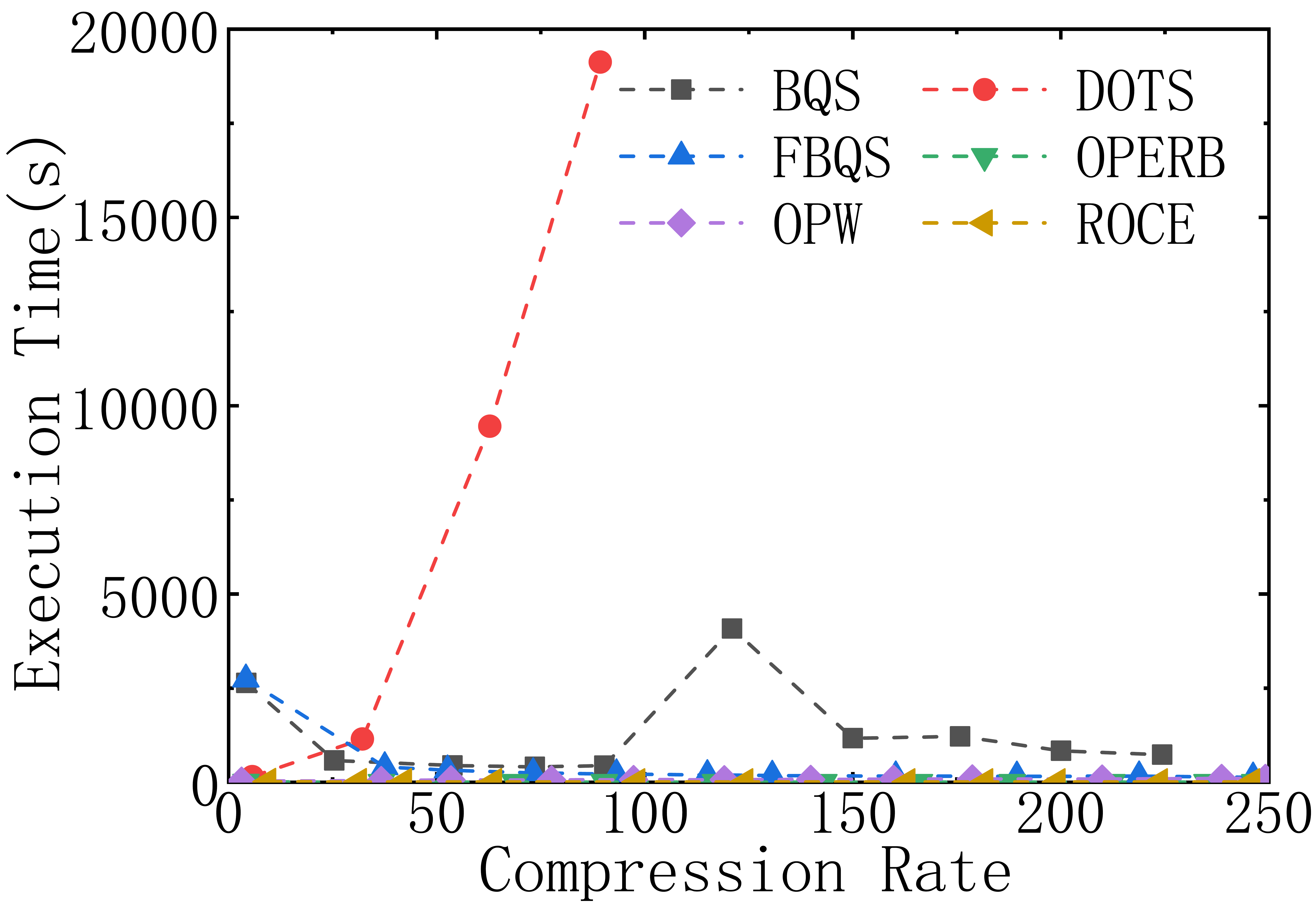}
		}
	\end{minipage}
	\begin{minipage}[t]{0.495\linewidth}
		\centering   
		\subfigure[Planet I]{ 
			\includegraphics[width=2.25in]{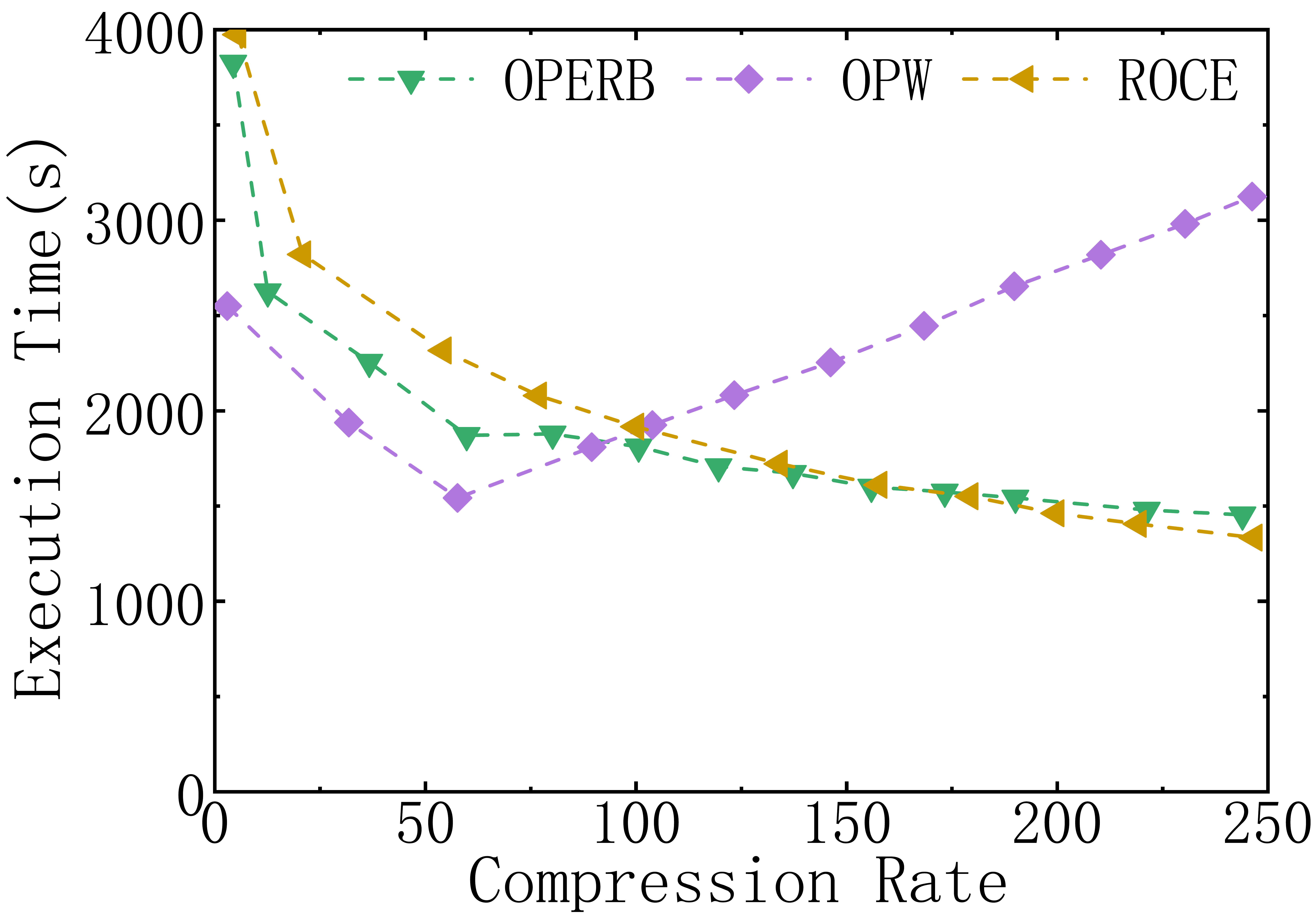}
		}
	\end{minipage}
	\caption{Efficiency evaluation: varying the compression rate}
	\label{figure:CompressionTime6}
\end{figure}

\subsubsection{Execution Time}\label{subsubsection:CompressionTime}
We first evaluate the execution times of these 6 algorithms w.r.t. varying the compression rate, and the results are reported in Figure \ref{figure:CompressionTime6}. ROCE, OPW and OPERB are obviously faster than BQS, FBQS and DOTS. During each loop iteration to compress a trajectory segment into a line segment, BQS and FBQS both need much more time than other algorithms at the beginning. Quite a bit of memory and time are needed by DOTS to handle the situations where the tracked object stays at the same place for a long time. And on Planet II, DOTS was too time-consuming to continue when the compression rate is close to 100, so we chose to stop the experiment. Only OPERB, OPW and ROCE were chosen to run on Planet I, since BQS, FBQS and DOTS are too slow to run on Planet I. On Planet I, the execution times of OPERB and ROCE are nearly the same, and both become shorter with the increase of the compression rate, because fewer trajectory segments need to be compressed into line segments. OPW needs more execution time with the increase of the compression rate, since the time complexity of OPW is $O(N^{2})$. And when the compression rate is more than 100, ROCE is faster than OPW. To conclude, ROCE is faster than many other compression algorithms.

\begin{figure}[ht]
	\centering
	\begin{minipage}[t]{0.495\linewidth}
		\centering
		\subfigure[Animal II]{
			\includegraphics[width=2.25in]{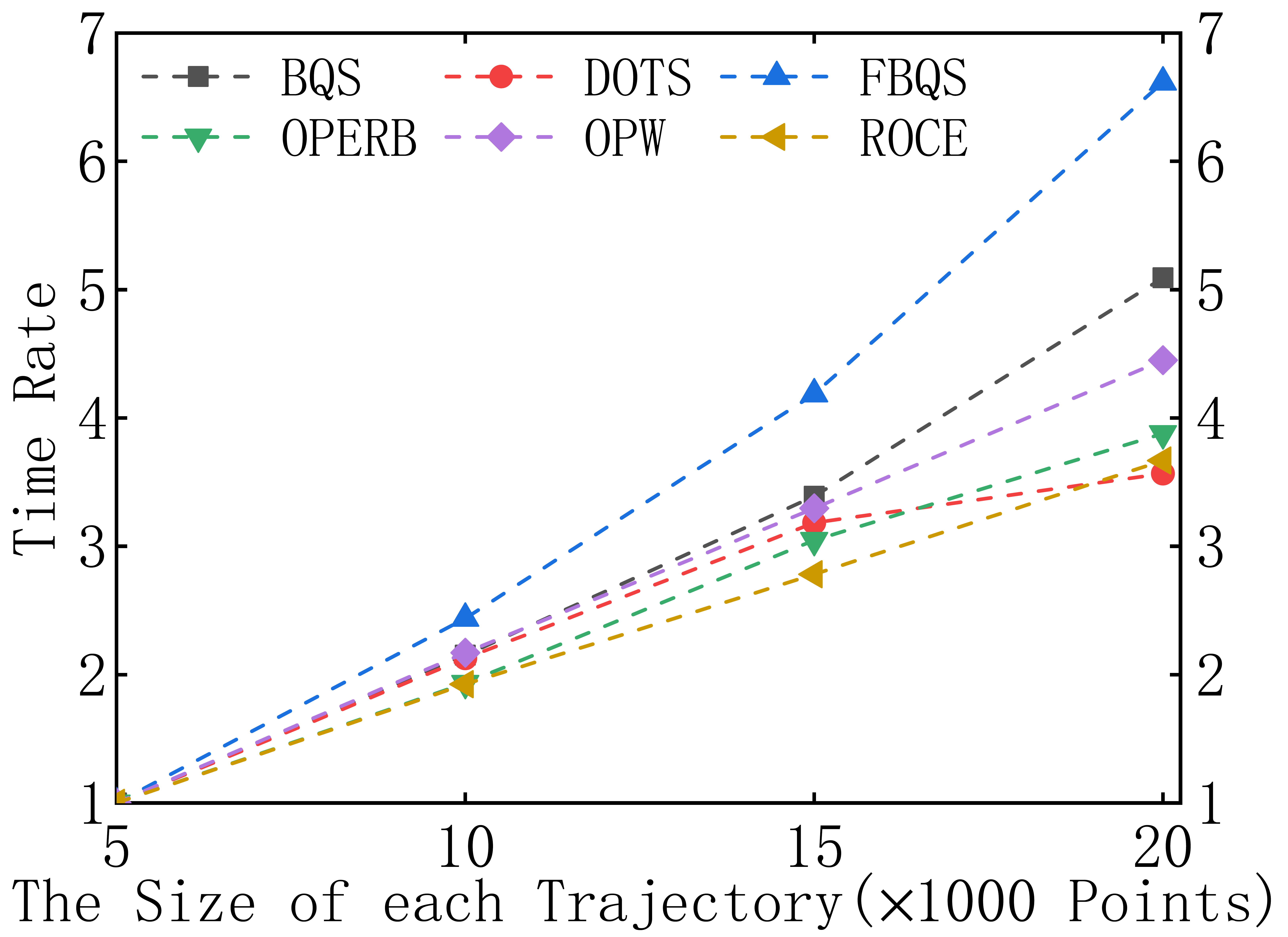}  
		}
	\end{minipage}    
	\begin{minipage}[t]{0.495\linewidth}
		\centering
		\subfigure[Indoor II]{
			\includegraphics[width=2.25in]{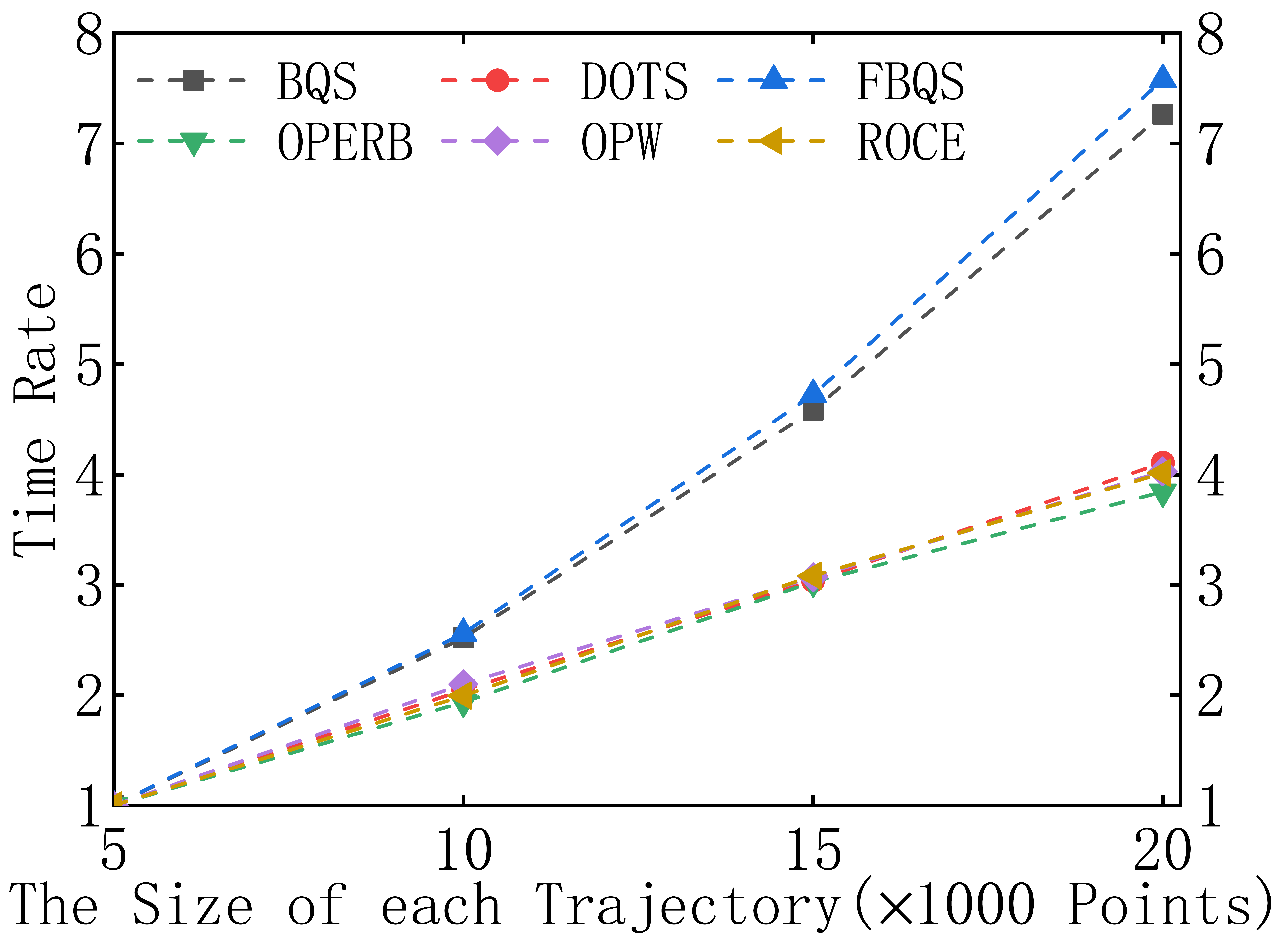}
		}
	\end{minipage}
	\begin{minipage}[t]{0.495\linewidth}
		\centering
		\subfigure[Planet II]{
			\includegraphics[width=2.25in]{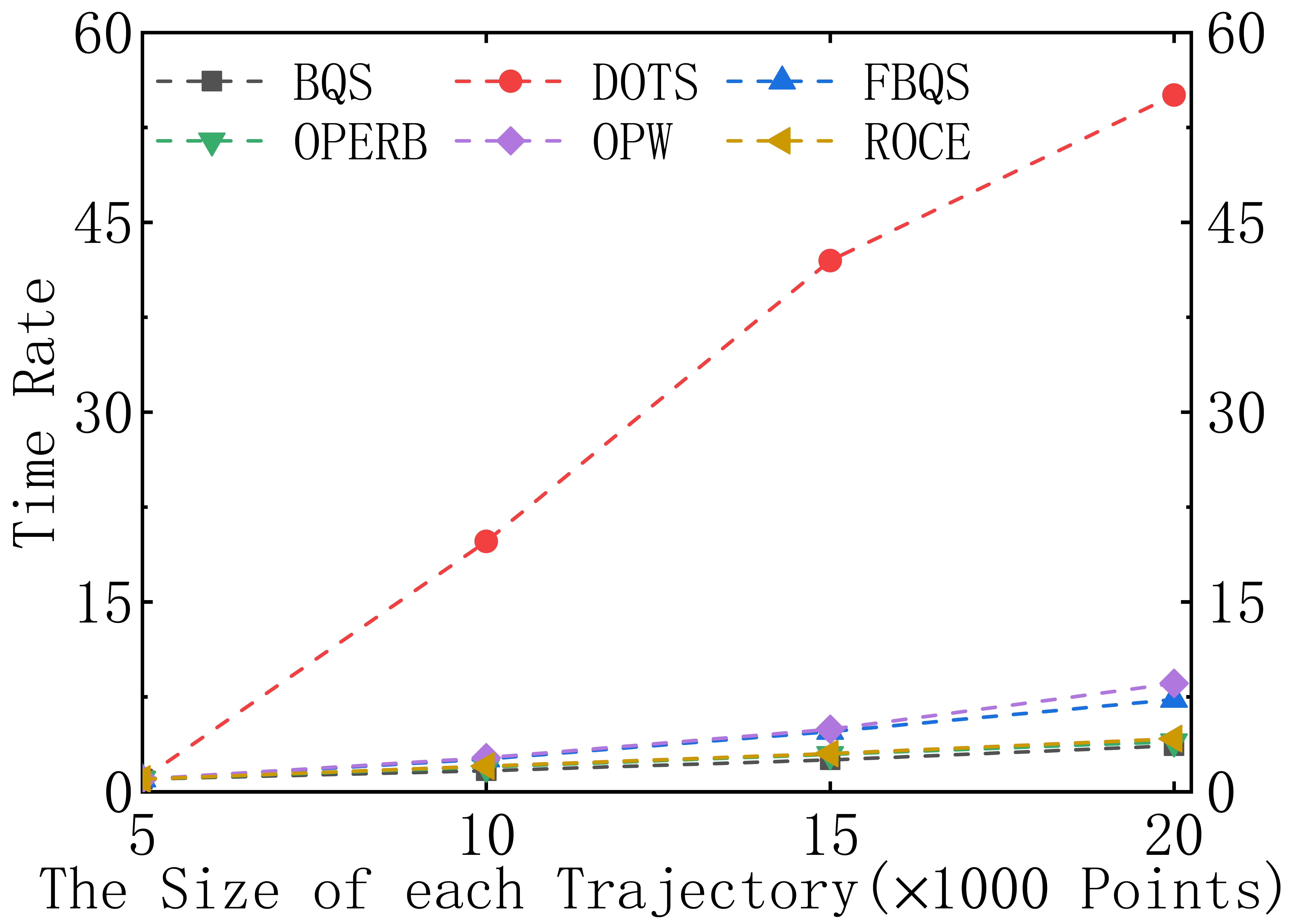}     
		}
	\end{minipage}  
	\caption{Efficiency evaluation: varying the size of trajectories}
	\label{figure:VaryTheLengthOfTrajectories}
\end{figure}

To evaluate the impact of the size of each raw trajectory (i.e. the number of points in each raw trajectory) on the execution time of compression, 20 trajectories with the largest sizes were chosen from Animal, Indoor and Planet  respectively, and the size of each trajectory was varied from 5000 to 20000 while the compression rates were all fixed as 50. The results are shown in Figure \ref{figure:VaryTheLengthOfTrajectories}. The y-coordinates are the time rates of the execution time of compression to the one of compressing trajectories whose sizes are all 5000. Only two algorithms ROCE and OPERB always scale well with the increase of the size of each trajectory on all datasets, and show nearly linear running time. But other algorithms do not, and much more execution time is needed to compress trajectories with larger sizes, especially for DOTS.

\begin{figure}[h]
	\centering
	\begin{minipage}{0.495\linewidth}
		\centering
		\subfigure[Animal II]{
			\label{figure:BoundedErrorOf6Algorithms.sub.1}
			\includegraphics[width=2.25in]{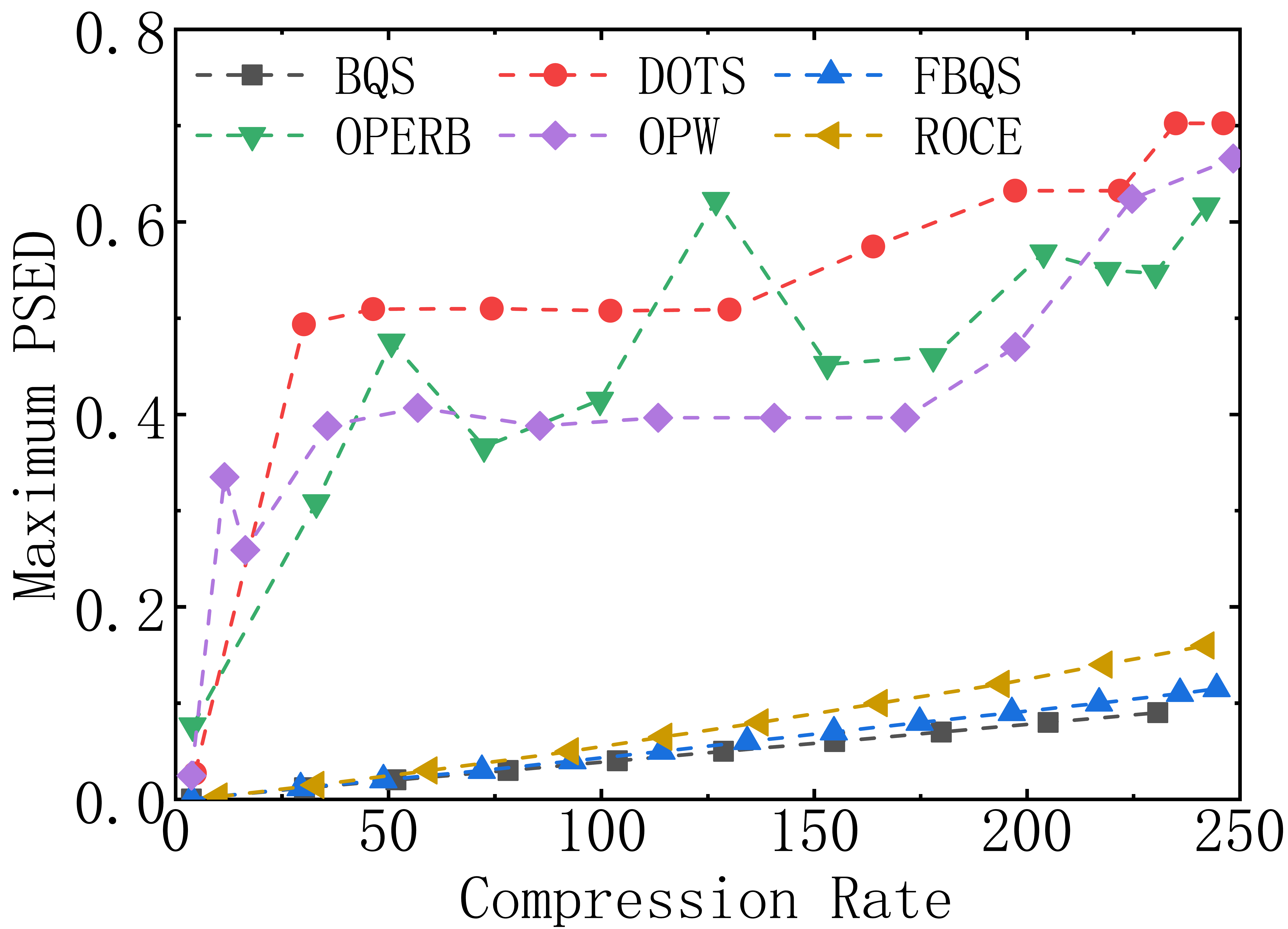}  
		} 
	\end{minipage}    
	\begin{minipage}{0.495\linewidth}
		\centering
		\subfigure[Indoor II]{
			\label{figure:BoundedErrorOf6Algorithms.sub.2}
			\includegraphics[width=2.25in]{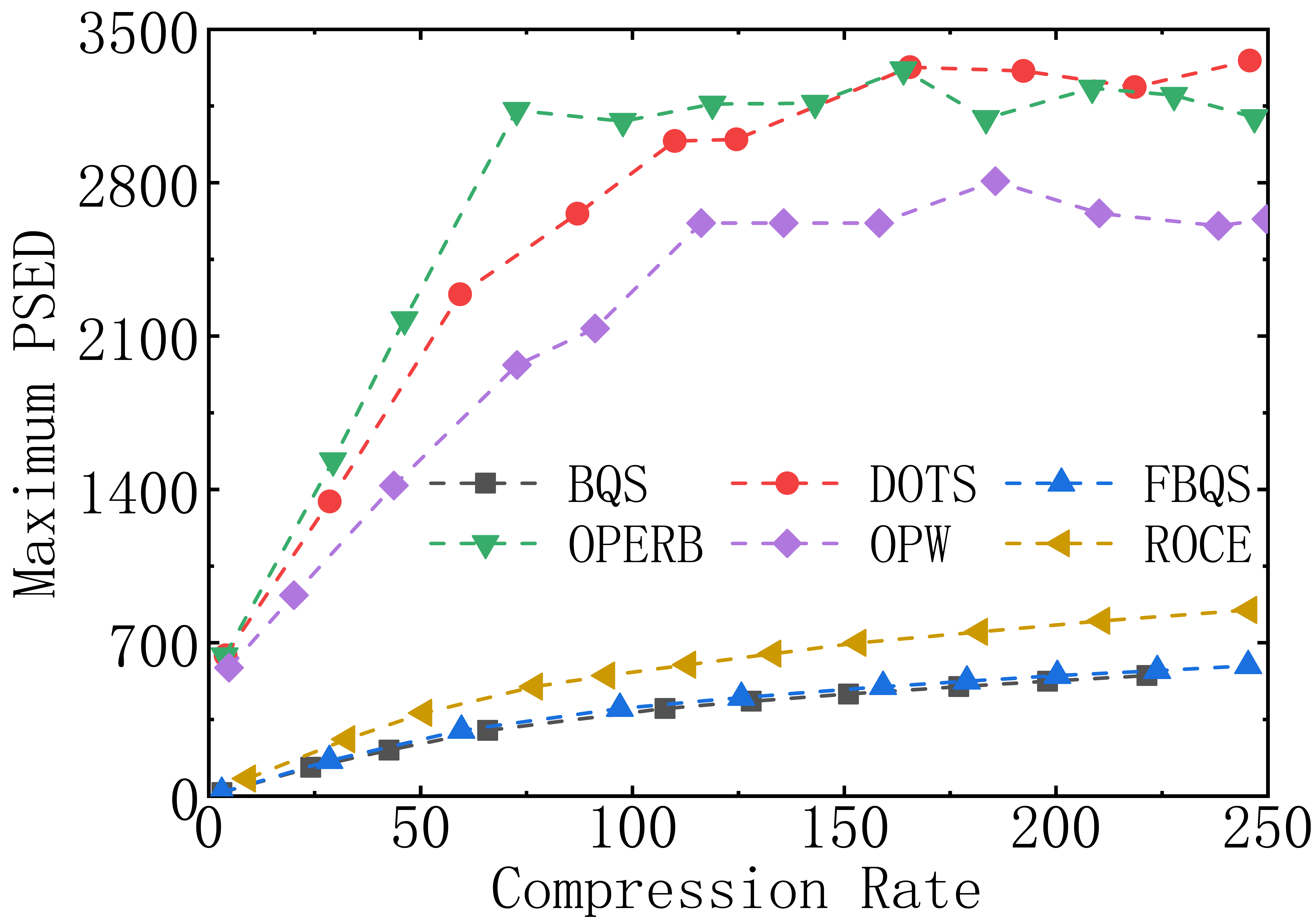}
		}
	\end{minipage}
	\begin{minipage}{0.495\linewidth}
		\centering
		\subfigure[Planet II]{
			\label{figure:BoundedErrorOf6Algorithms.sub.3}
			\includegraphics[width=2.25in]{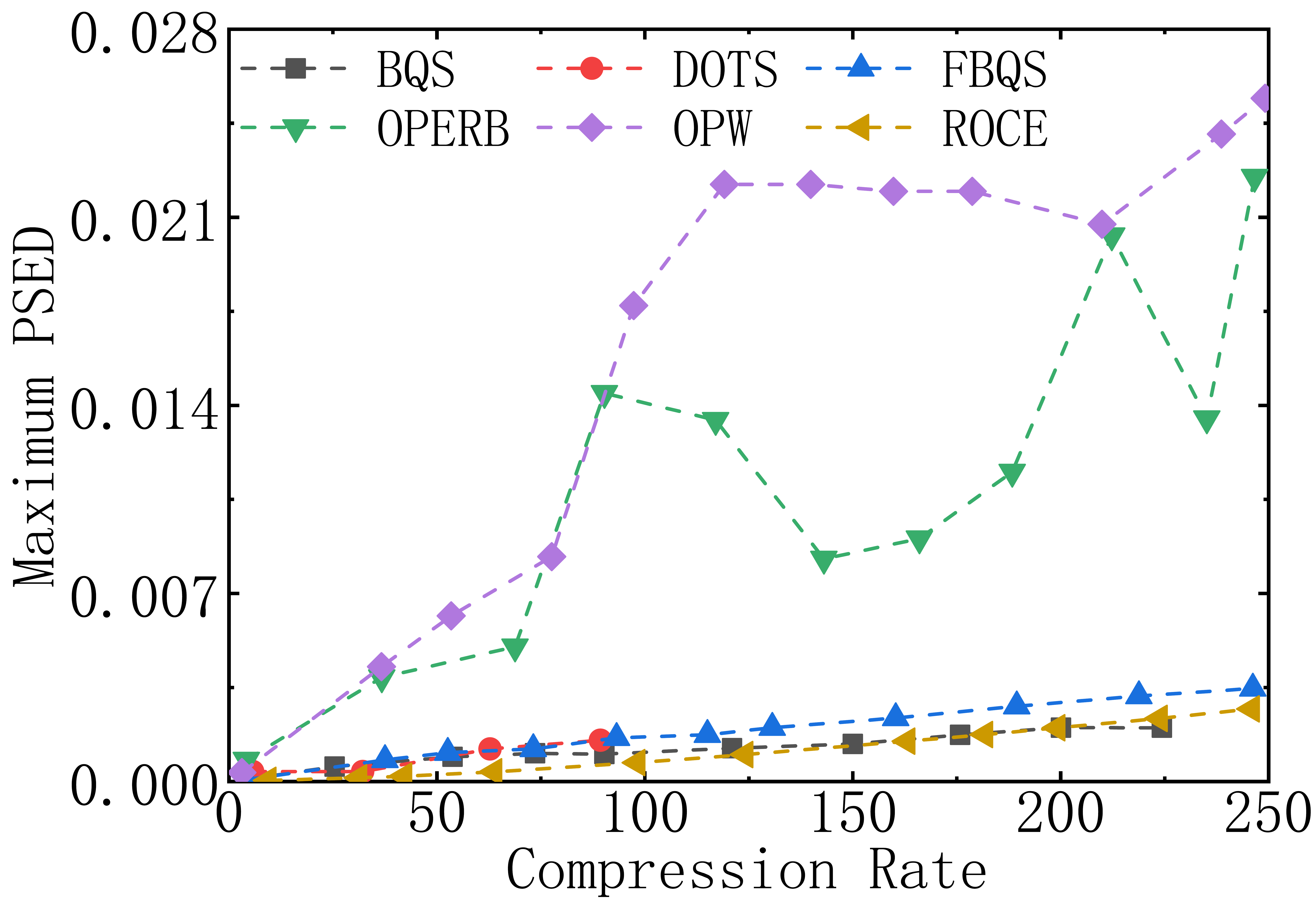}     
		}
	\end{minipage}
	\begin{minipage}{0.495\linewidth}
		\centering
		\subfigure[Animal II]{
			\includegraphics[width=2.25in]{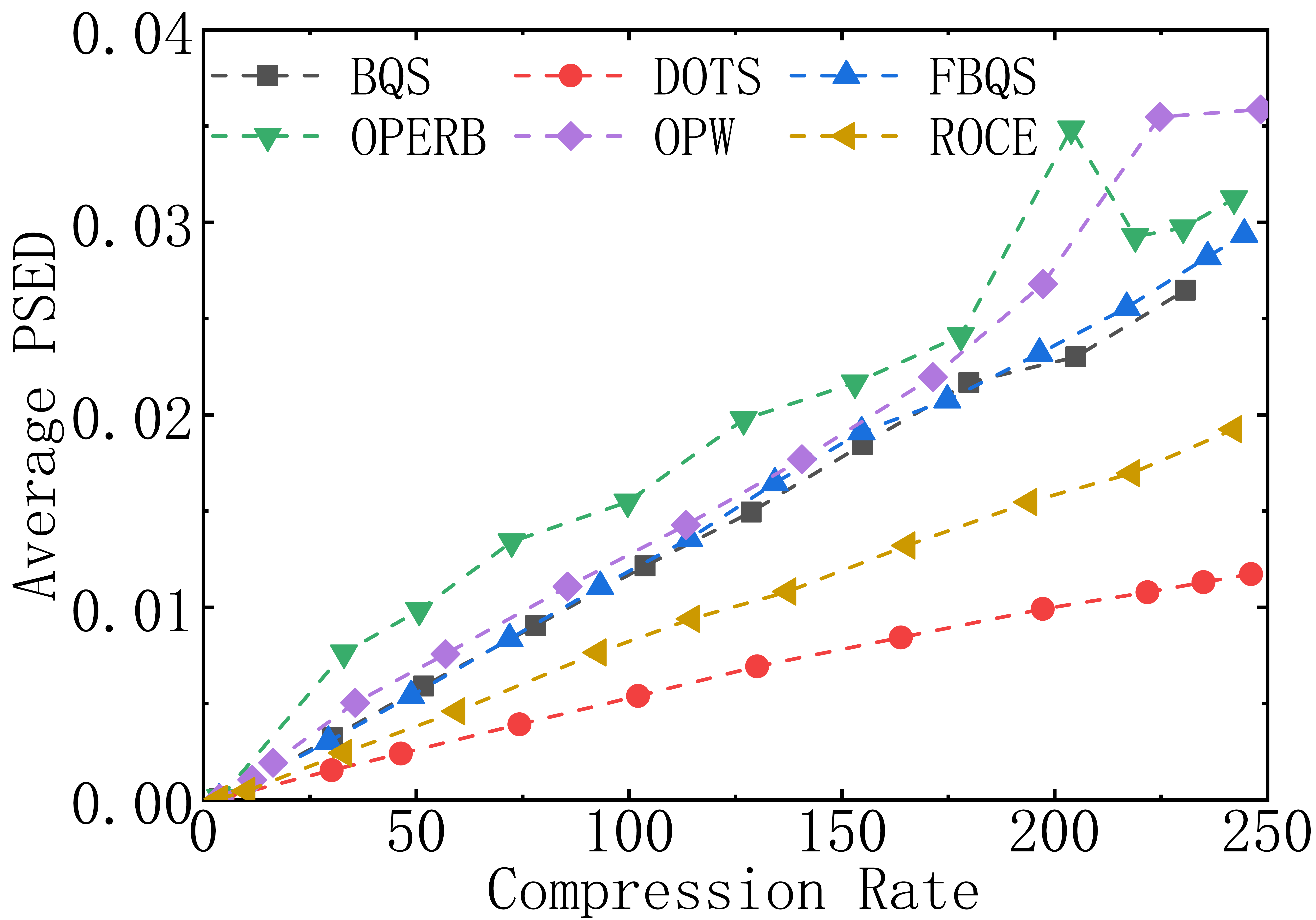}  
		}
	\end{minipage}
	\begin{minipage}{0.495\linewidth}
		\centering   
		\subfigure[Indoor II]{
			\includegraphics[width=2.25in]{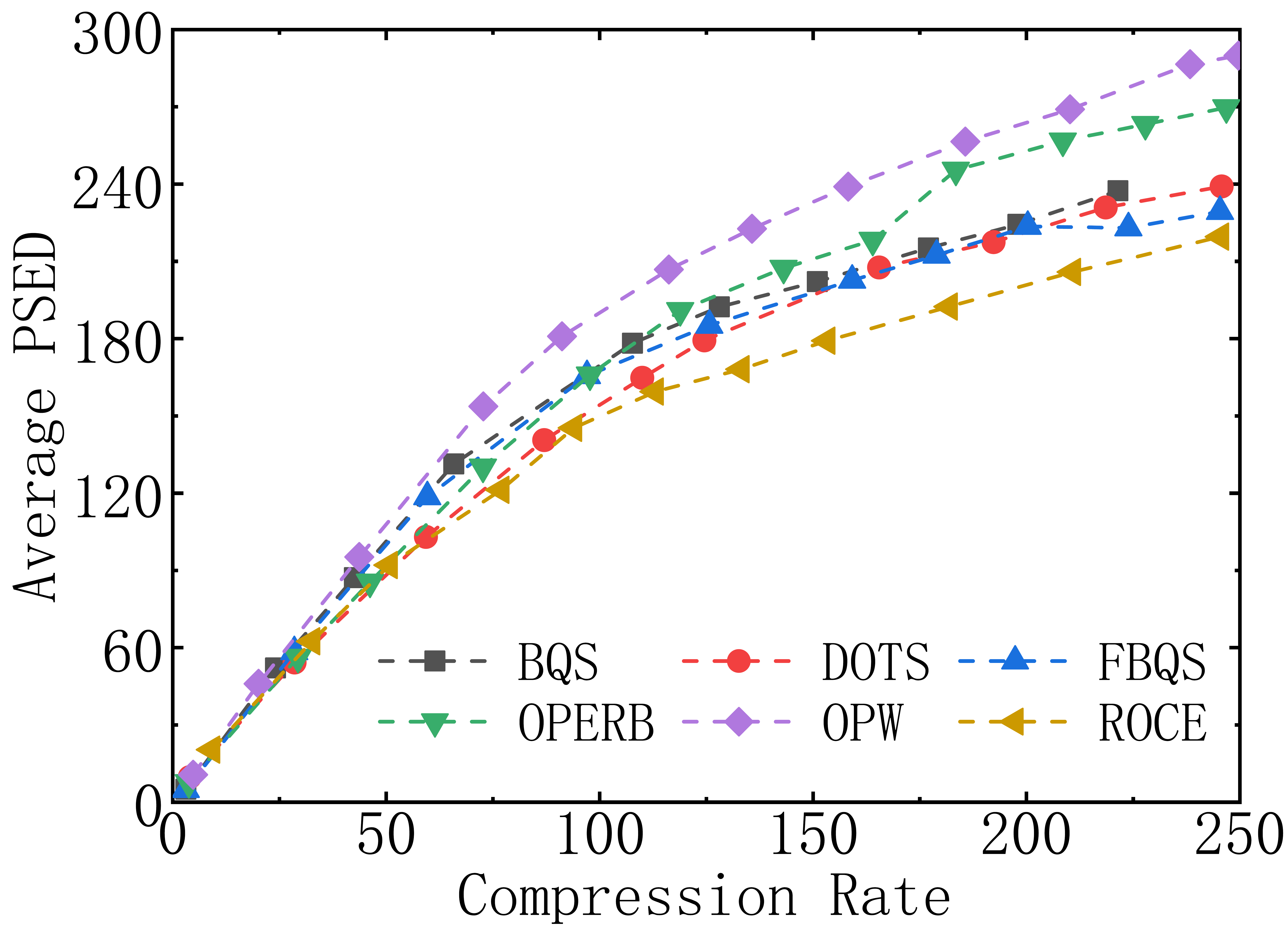}
		}
	\end{minipage}
	\begin{minipage}{0.495\linewidth}
		\centering
		\subfigure[Planet II]{ 
			\includegraphics[width=2.35in]{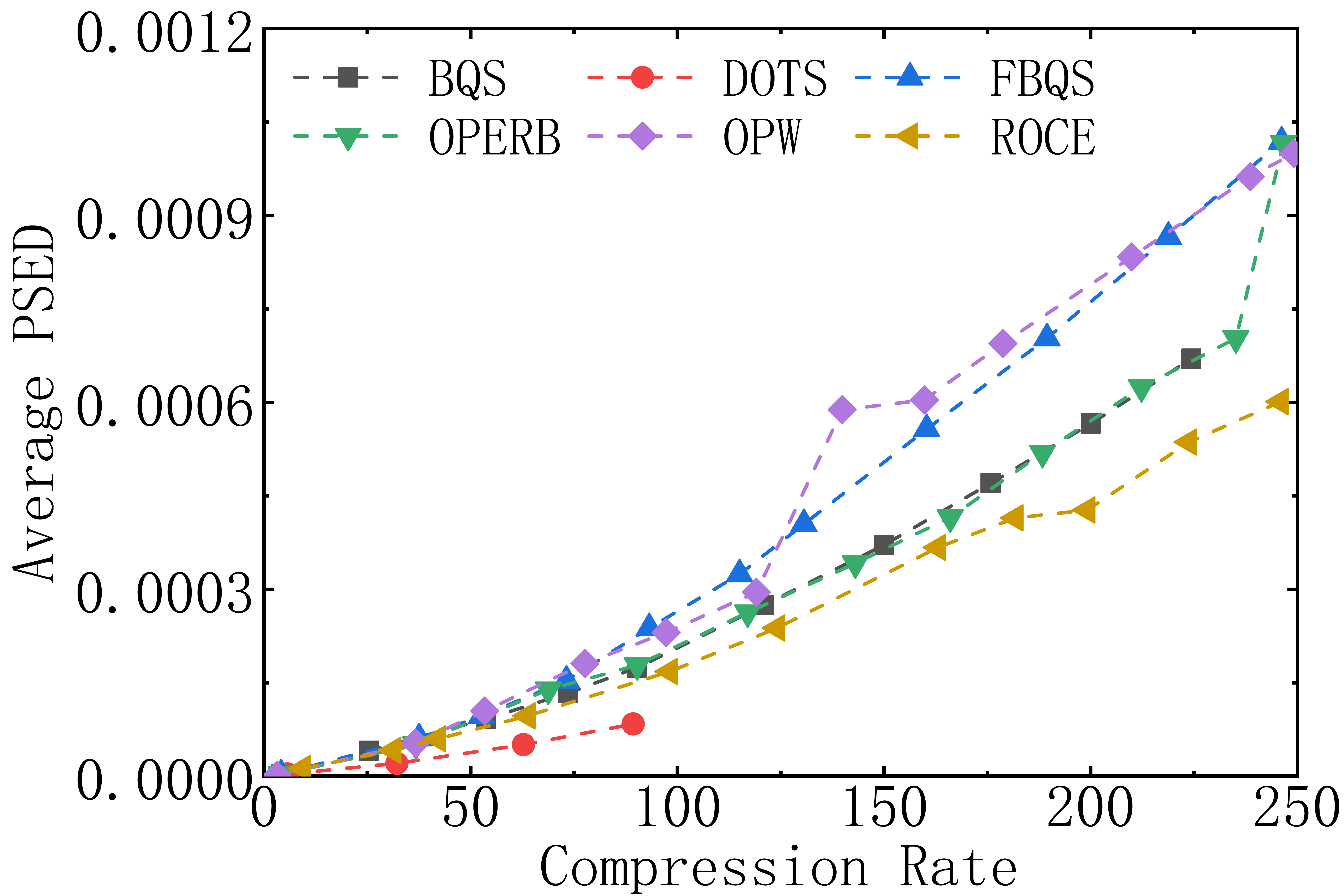}     
		}    
	\end{minipage}
	\caption{Evaluation of the maximum and the average PSED: varying the compression rate}
	\label{figure:BoundedErrorOf6Algorithms}
\end{figure}

\subsubsection{Accuracy Loss}\label{subsubsection:AccuracyLoss}

For these compression algorithms, in order to compare the accuracy loss of the generated compressed trajectories the maximum and the average PSED of these compressed trajectories are evaluated w.r.t. varying the compression rate, and the results are reported in Figure \ref{figure:BoundedErrorOf6Algorithms}. For all algorithms, both the maximum and the average PSED increase with the increase of the compression rate. In terms of execution time, ROCE performs similarly with OPERB and OPW, but ROCE always performs much better than OPERB and OPW on the maximum PSED. ROCE, BQS and FBQS always perform similarly on the maximum PSED, but both BQS and FBQS need much more execution time than ROCE. ROCE always performs much better than most other algorithms on the average PSED. So with much less exection time, the compressed trajectories generated by ROCE maintain much less accuracy loss than those generated by most other algorithms. So, it is quite clear that ROCE makes the best balance among the accuracy loss, the time cost and the compression rate. Among the fastest algorithms, the accuracy loss of the compressed trajectories generated by ROCE is always the smallest, and among algorithms with the smallest accuracy loss, ROCE is always the fastest.

\subsection{Performance Evaluation for RQC Algorithm}\label{subsection:RangeQueryProcessingAlgorithmRQC}

The range query serves as a primitive, yet essential operation. In the previous work related to range queries, such as \cite{zhang2018trajectory,zhang2018gpu,dong2018gat}, each trajectory is usually seen as a sequence of discrete points, and a trajectory is regarded to be overlapped with the query region $R$ iff at least one point in this trajectory falls in $R$. However, this traditional criteria is completely unsuitable for range quering on compressed trajectories, and many trajectories will be missing in the result set. To solve this problem, we propose the new criteria defined in Definition \ref{definition:RangeQuery} and the efficient RQC algorithm for range queries on compressed trajectories. On compressed trajectories, whether the traditional or the new criteria is used results in quite different range query results. In the first following experiment, we evaluate the deviation between the range query results on raw trajectories and the corresponding compressed trajectories based on different criteria. The impacts of the size of each query region, the number $n_{s}$ of sampling points, the probality threshold value $p$, \emph{ASP\_tree} are also respectively studied.

\subsubsection{Range Queries on Compressed Trajectories Based on the Traditional or the New Criteria}\label{subsubsection:RangeQuerybasedonSegments}

To measure the deviation between the range query results on the raw trajectories and those on the corresponding compressed trajectories, 3 evaluation metrics are defined and used here. Given a range query, the query result set on the raw trajectories based on the traditional criteria is denoted by $Q_{R}$. And $Q_{C}$ represents the query result set on compressed trajectories. The precision rate $Pre$ and the recall rate $Rec$ of a range query result on compressed trajectories are respectively defined as: \begin{displaymath}Pre=\frac{|Q_{R} \cap Q_{C}|}{|Q_{C}|} \ \ \ \ \ \ \ \ \ \ \  \ Rec=\frac{|Q_{R} \cap Q_{C}|}{|Q_{R}|}\end{displaymath} $F_{1}$-Measure, a comprehensive evaluation metric, is defined as:\begin{displaymath}\frac{1}{F_{1}}=\frac{1}{2}*(\frac{1}{Pre}+\frac{1}{Rec})\end{displaymath}

\begin{figure*}[ht]
	\centering
	\begin{minipage}[t]{0.495\linewidth}
		\centering
		\subfigure{
			\includegraphics[width=2.25in]{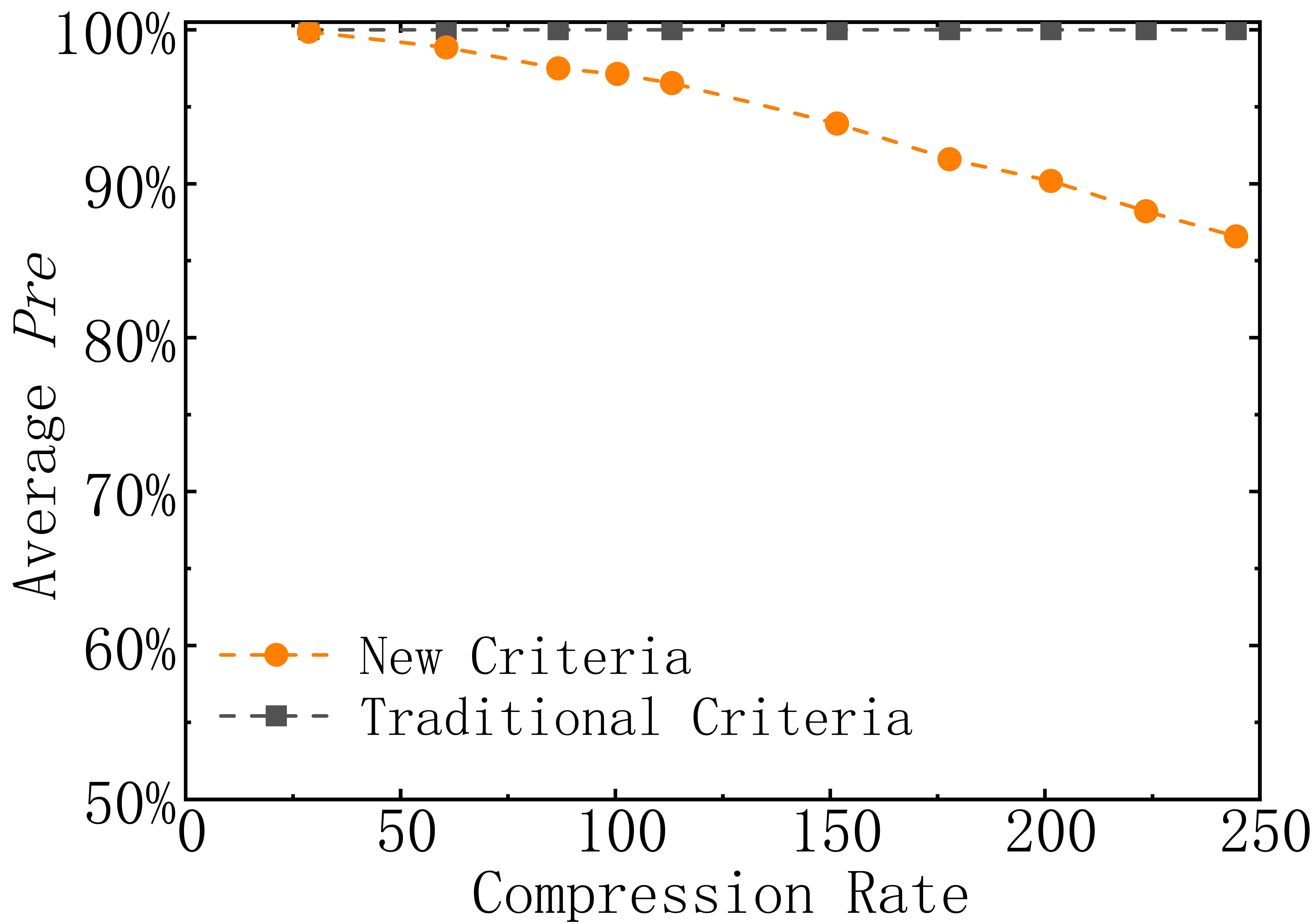}
		}
	\end{minipage}
	\begin{minipage}[t]{0.495\linewidth}
		\centering
		\subfigure{
			\includegraphics[width=2.25in]{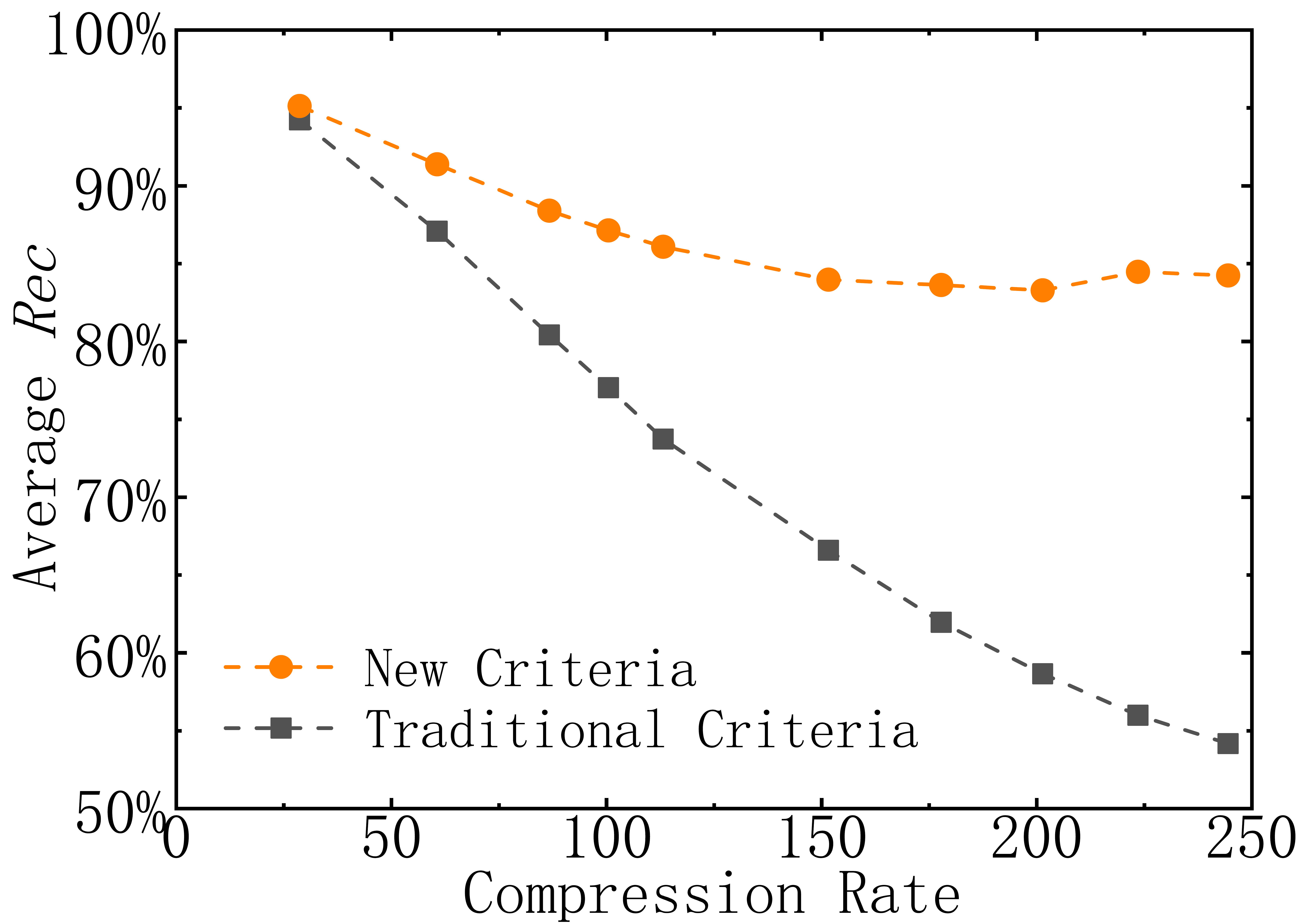}
		}
	\end{minipage}
	\begin{minipage}[t]{0.495\linewidth}
		\centering
		\subfigure{ 
			\includegraphics[width=2.25in]{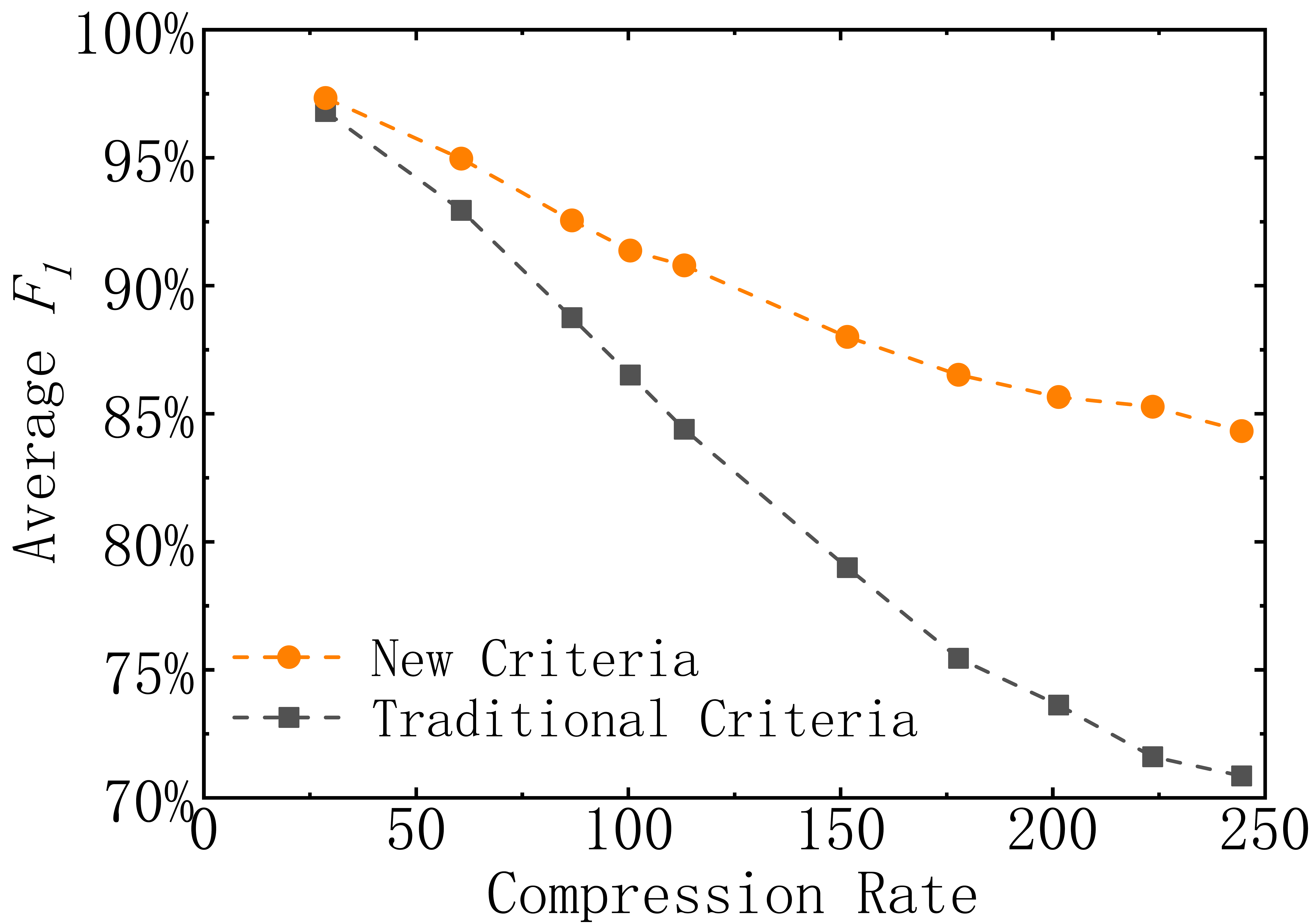} 
		}
	\end{minipage}
	\caption{Evaluation of the average $Pre$, $Rec$ and $F_{1}$ of range query results on compressed trajectories: varying the compression rate}
	\label{figure:RangeQueryBasedOnPointVSSegment}
\end{figure*}

For 1000 randomly generated range queries on compressed trajectories, we evaluate the average $Pre$, $Rec$ and $F_{1}$ of the query results w.r.t. varying the compression rate, and the results are shown in Figure \ref{figure:RangeQueryBasedOnPointVSSegment}. When querying on compressed trajectories based on the traditional criteria, the average $Pre$ is always 1, since all points in each compressed trajectory must be in the corresponding raw trajectors. But with the increase of the compression rate, $Rec$ declines sharply, which shows that up to 45.8\% of the corresponding raw trajectories overlapped with the query regions are not discovered. When querying on compressed trajectories based on the new criteria, though at most 13.4\% of the corresponding raw trajectories not overlapped with the query regions appear in the result set, much more corresponding raw trajectories (i.e. at least 84.2\%) overlapped with the query regions can be found out. The average $F_{1}$, the comprehensive metric of $Pre$ and $Rec$, also demonstrates the stable superiority of the new criteria. In summary, it is far more suitable to answer range queries on compressed trajectories based on the new criteria.

\subsubsection{Impacts of the Sizes of Query Regions}\label{EffectofQueryRegion} 
\begin{figure*}[ht]
	\centering
	\begin{minipage}[t]{0.48\linewidth}
		\centering
		\includegraphics[width=2.25in]{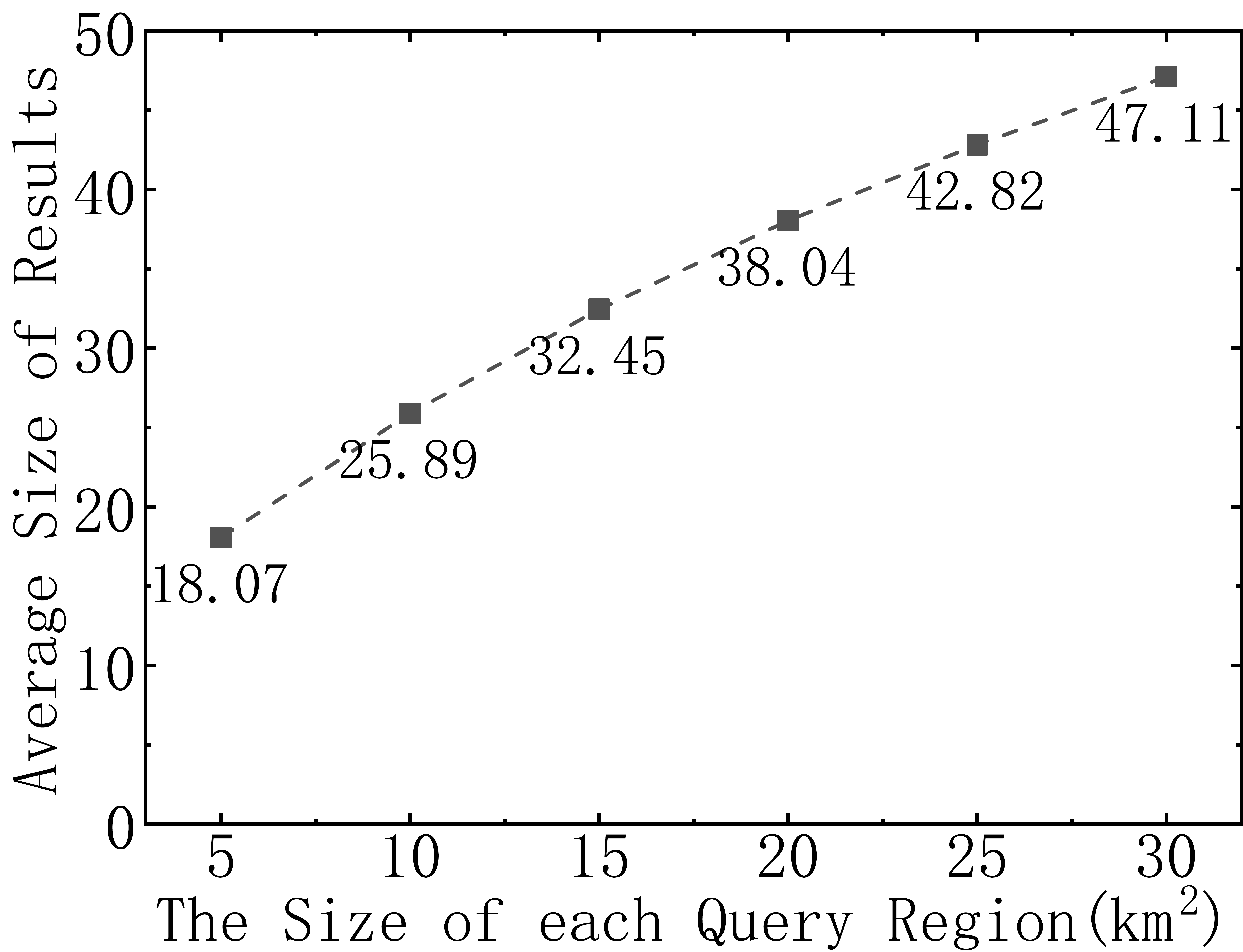}
		\caption{Evaluation of the average size of range query results on compressed trajectories: varying the size of each query region}
		\label{figure:VaryTheAreaOfQueryRegionsSeeResult}
	\end{minipage}\ \ \ \ \
	\begin{minipage}[t]{0.48\linewidth}
		\centering
		\includegraphics[width=2.25in]{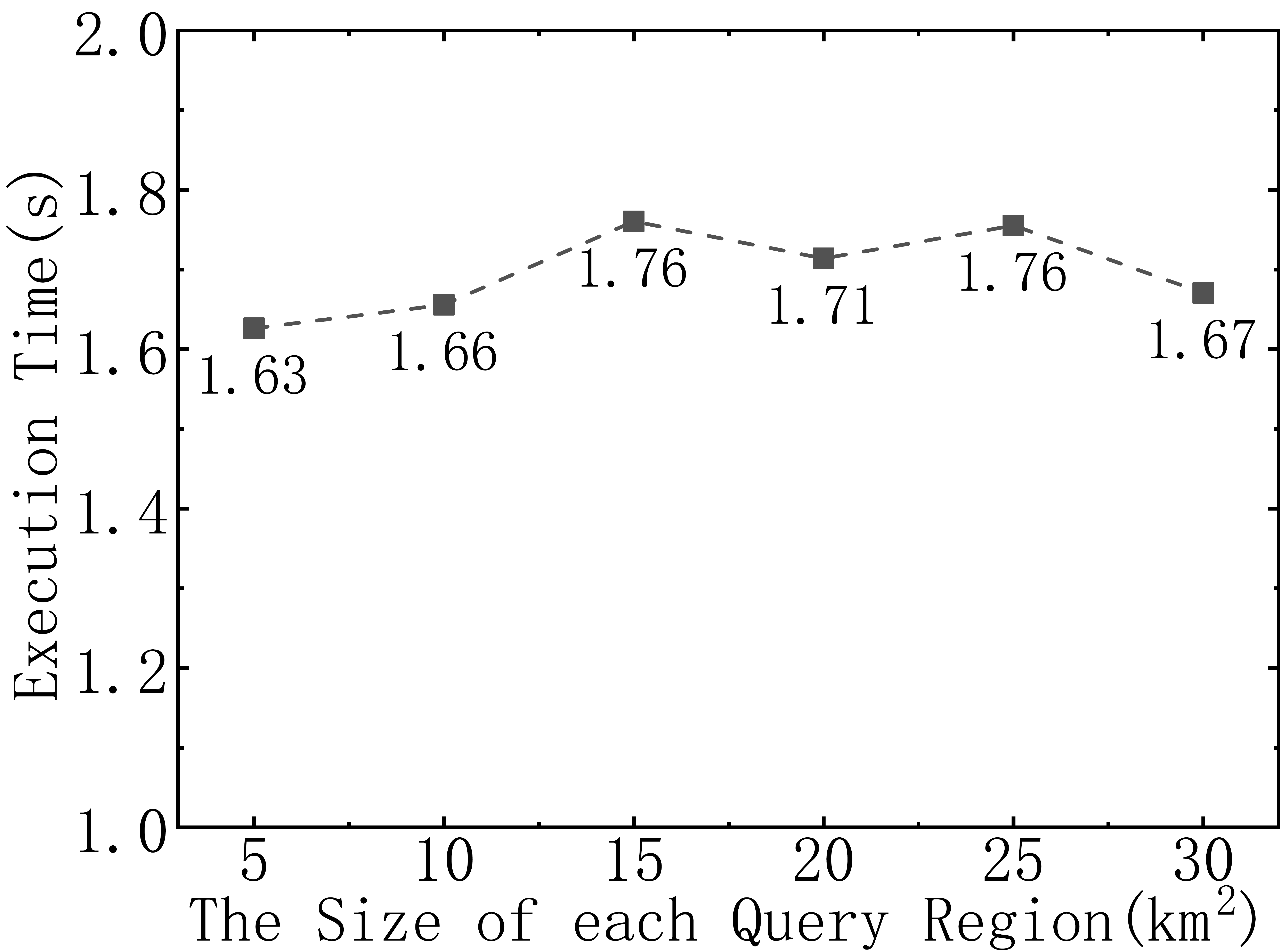}
		\caption{Efficiency evaluation: varying the size of each query region}
		\label{figure:VaryTheAreaOfQueryRegions}
	\end{minipage}
\end{figure*}

For range queries, the size of the query regions may have impact on the size of range query results and the execution time. 1000 randomly generated range queries were executed on the compressed trajectories whose compression rate is 200, and the size of each query region was varied from 5$km^{2}$ to 30$km^{2}$ in area. The results are shown in Figure \ref{figure:VaryTheAreaOfQueryRegionsSeeResult} and Figure \ref{figure:VaryTheAreaOfQueryRegions}. We can see that with the increase of the size of each query region, though the average size of query results on compressed trajectories grows nearly linearly, the execution time needed by our RQC algorithm does not vary much. Thus RQC algorithm can easily support range queries with much larger query regions without needing more execution time. And RQC algorithm is quite efficient and a range query on more than 90000 compressed trajectories can be processed just within 2ms.

\subsubsection{Impacts of the Number $n_{s}$ of Sampling Points}\label{Effectofns}

\begin{figure}[ht]
	\centering
	\begin{minipage}[t]{0.48\linewidth}
		\centering
		\includegraphics[width=2.35in]{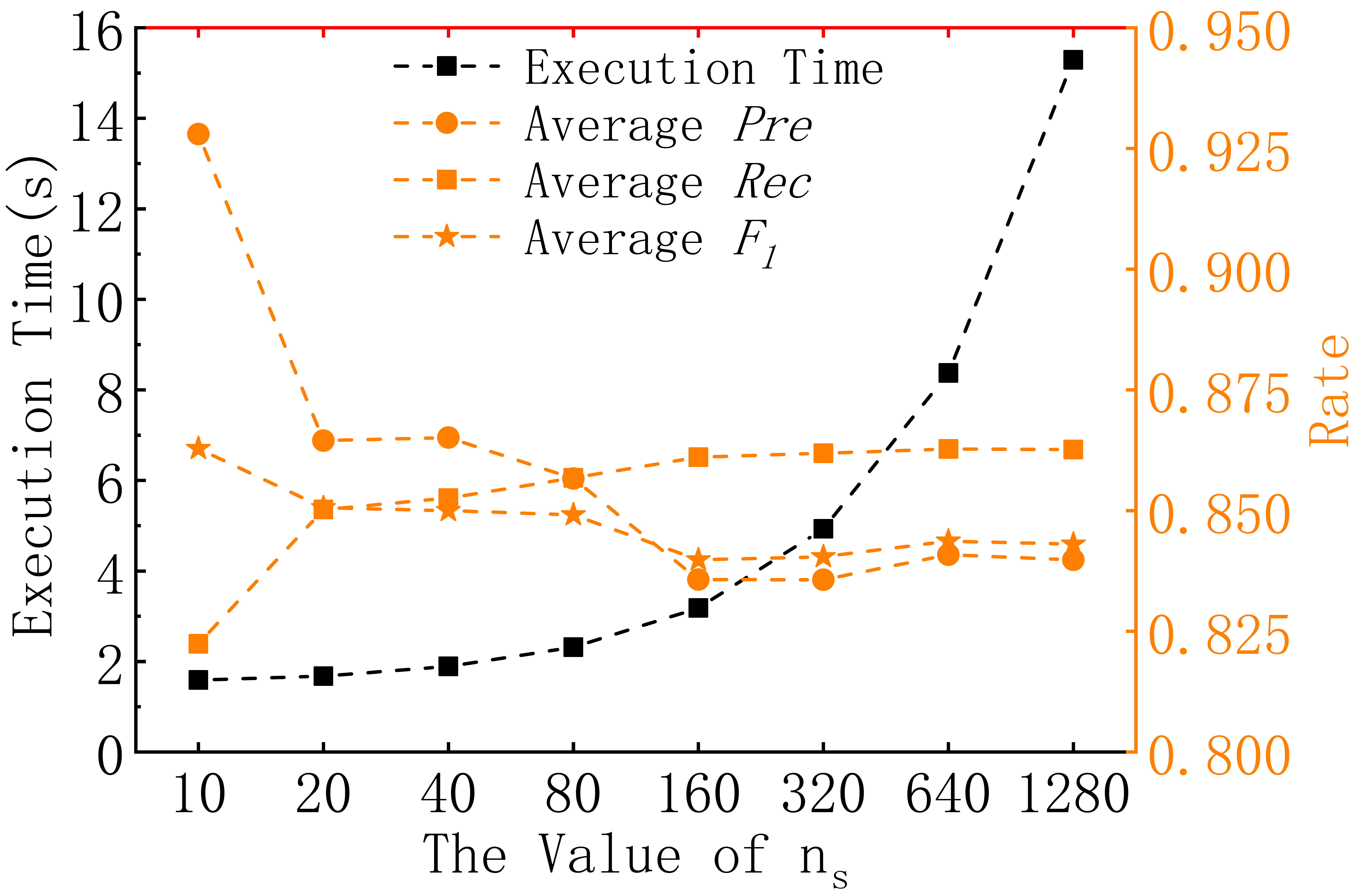}
		\caption{Evaluation of the execution time, the average $Pre$, $Rec$ and $F_{1}$: varying the value of $n_{s}$}
		\label{figure:VaryNs}
	\end{minipage}\ \ \ \ \
	\begin{minipage}[t]{0.48\linewidth}
		\centering
		\includegraphics[width=2.2in]{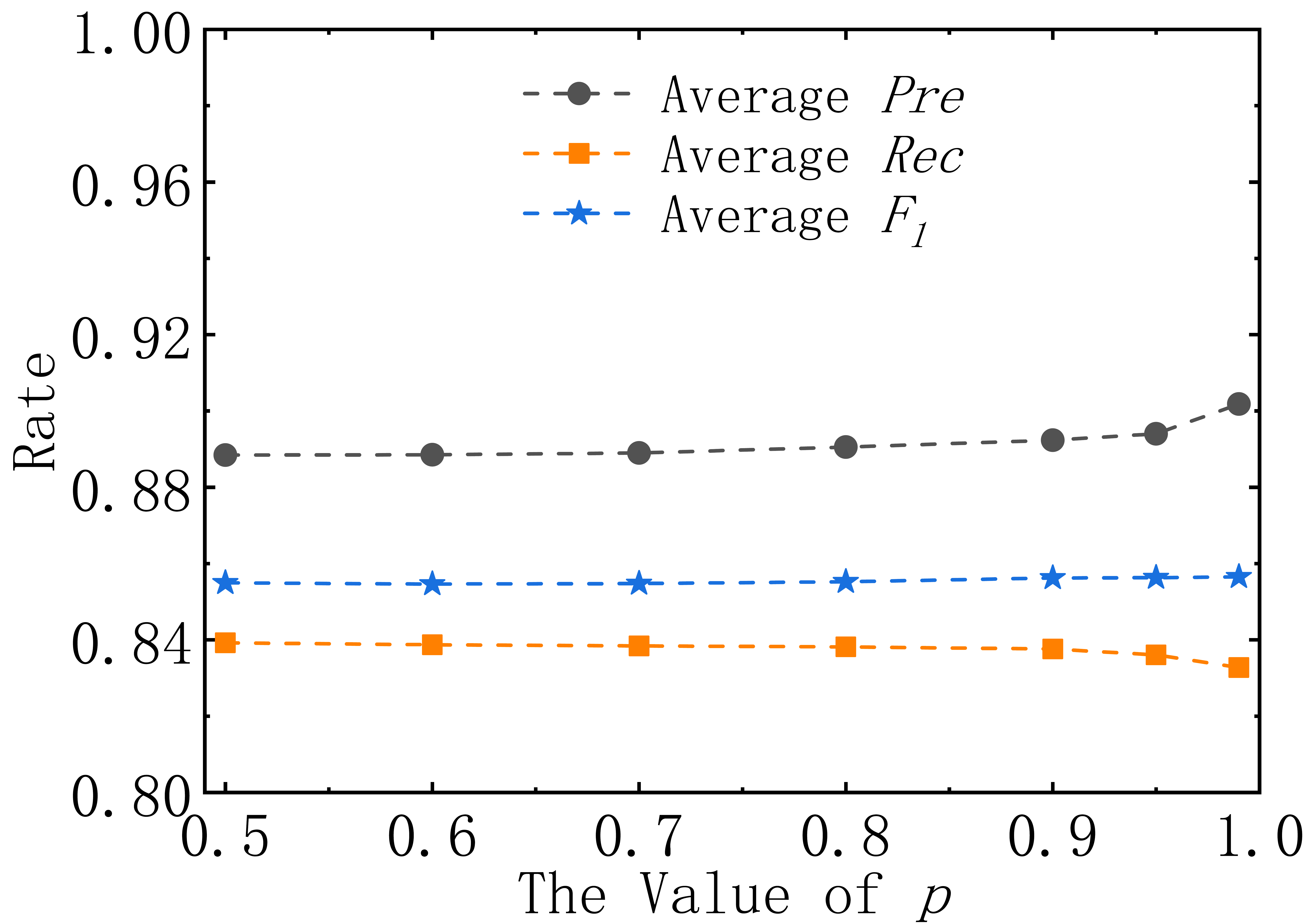}
		\caption{Evaluation of the average $Pre$, $Rec$ and $F_{1}$: varying the value of $p$}
		\label{figure:Varyp}
	\end{minipage}
\end{figure}

To process a range query on compressed trajectories, $n_{s}$ points are sampled to calculate the approximate probablity of that there exists a discarded point of its corresponding line segment falling in the query region $R$. We evaluate the execution time, the average precision rate $Pre$, the average recall rate $Rec$ and the average $F_{1}$ of query results of 1000 randomly generated range queries on the compressed trajectories whose compression rate is 200 w.r.t. varying $n_{s}$ from 10 to 1280, and the results are reported in Figure \ref{figure:VaryNs}. As $n_{s}$ increases, which means that more points are sampled to calculate the approximate probablity, obviously more execution time is needed. On one hand, the average $Pre$ gets a little smaller since a little more compressed trajectories, whose corresponding raw trajectories are not overlapped with the query region, are determined to be in the range query result set. On the other hand, the average $Rec$ gets a little larger because a little more compressed trajectories, whose corresponding raw trajectories are overlapped with the query region, can be found by our RQC algorithm. And the average $F_{1}$, a comprehensive evaluation metric, does not change much. In order to make a good balance between the execution time and the quality of range query result on compressed trajectories, $n_{s}$ is set to 15 as the default value in other experiments.

\subsubsection{Impacts of the Probability Threshold Value $p$}\label{Effectofp}

For a compressed trajectory, only when the calculated probability of its corresponding raw trajectory overlapped with query region $R$ is larger than $p$, can this compressed trajectory be in the range query result set. We evaluate the average precision rate $Pre$, the average recall rate $Rec$ and the average $F_{1}$ of query results of 1000 randomly generated range queries on the compressed trajectories whose compression rate is 200 w.r.t. varying $p$ from 0.5 to 0.99, and the results are shown in Figure \ref{figure:Varyp}. With the increase of $p$, we can see that for the average precision rate $Pre$ and the average recall rate $Rec$, especially the average $F_{1}$, they do not change much, which shows that changing the value of $p$ has quite limited influence on the quality of range query results on compressed trajectories.

\subsubsection{Impacts of ASP\_tree}\label{EffectofASP-tree}
\begin{figure}[ht]
	\centering
	\begin{minipage}[t]{0.48\linewidth}
		\centering
		\includegraphics[width=2.25in]{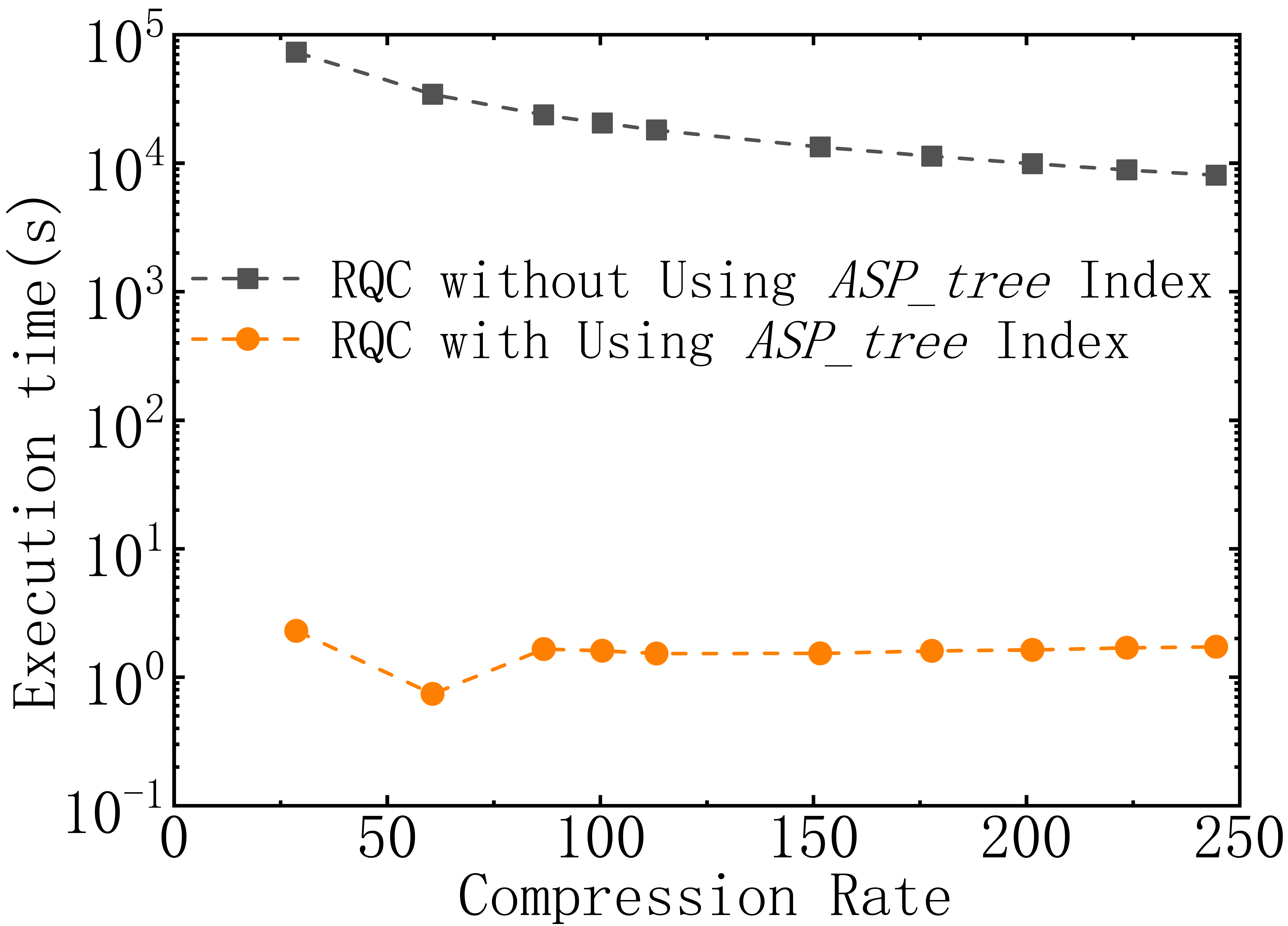}
		\caption{Efficiency evaluation: \emph{ASP\_tree} index when varying the compression rate}
		\label{figure:TimeNeededWithAndWithoutASP-tree}
	\end{minipage}\ \ \ \ \
	\begin{minipage}[t]{0.48\linewidth}
		\centering
		\includegraphics[width=2.35in]{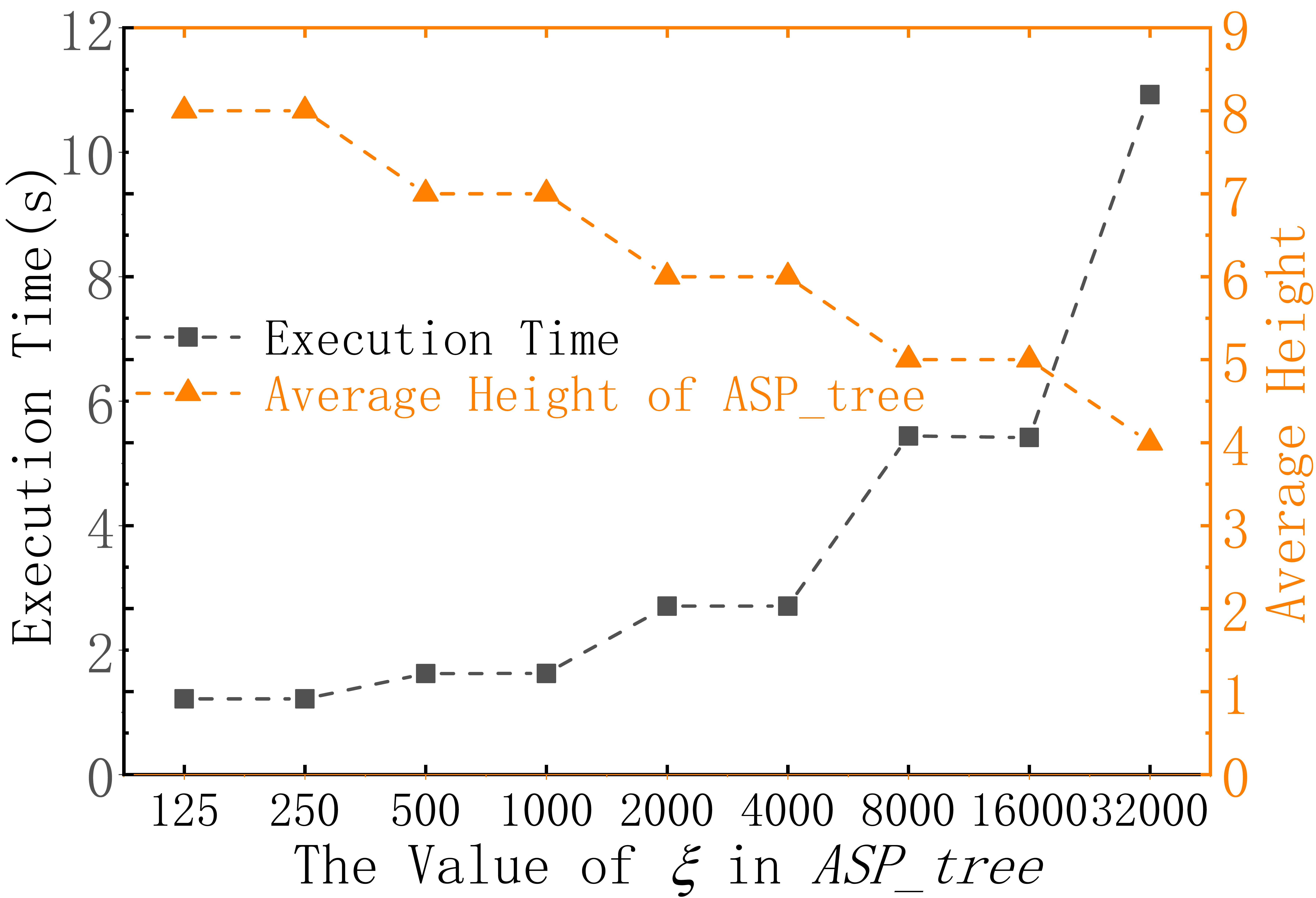}
		\caption{Evaluation of the execution time and the average height of \emph{ASP\_tree}: varying the value of $\xi$ in \emph{ASP\_tree}}
		\label{figure:AverageHeightAndRangeQueryTime}
	\end{minipage}
\end{figure}

How much \emph{ASP\_tree} index can accelerate the range query processing is studied first. The execution time of 1000 randomly generated range queries is evaluated w.r.t. varying the compression rate, and the results are reported in Figure \ref{figure:TimeNeededWithAndWithoutASP-tree}. By using \emph{ASP\_tree}, the range query processing can be accelerated extremely. At least 99.98\% of the execution time can be easily saved, and the acceleration gets more obvious with the decrease of the compression rate of the queried trajectories.

For each node in \emph{ASP\_tree}, if there are more than $\xi$ endpoints of all line segments falling in the corresponding region of this node, then this node is a non-leaf node with 4 child nodes. So the threshold value $\xi$ controls the average height of \emph{ASP\_tree}. The average height of \emph{ASP\_tree} is defined as the average height of all leaf nodes, and the height of the root node is defined as 1. We evaluate the average height of \emph{ASP\_tree} and the execution time of 1000 randomly generated range queries on the compressed trajectories whose compression rate is 200 w.r.t. varying $\xi$, and the results are reported in Figure \ref{figure:AverageHeightAndRangeQueryTime}. The heights of all leaf nodes are always the same, which means that \emph{ASP\_tree} is always well balanced, our region partitioning strategy dose work, and data skew is perfectly avoided. We can see that $\xi$ controls the average height of \emph{ASP\_tree}, and the average height declines with the increase of $\xi$. The average height has direct impact on the execution time of range queries. Less execution time is needed by these range queries with the increase of the average height of \emph{ASP\_tree}, because much more compressed trajectories not in the final result set are directly filtered out by searching \emph{ASP\_tree}.

\section{Related Work}\label{section:RelatedWork}
$\mathbf{Trajectory \ Compression \ algorithms \ in \ online \ mode.}$ Able to be applied in much more application scenarios, some trajectory compression algorithms in online mode based on different accuracy loss metrics have been proposed and attract people's attention. PED, SED, DAD and LISSED are 4 frequently-used accuracy loss metrics, which measure the degree of the accuracy loss after a trajectory is compressed. There is no clear evidence that there exists an accuracy loss metric superior to all the others in the literature. On one hand, DAD, a direction-based distance, is defined based on the greatest angular difference between two directions. Since DAD does not provide any error guarantee on the distance, for the compressed trajectories generated by compression algorithms based on DAD\cite{long2013direction,2014Trajectory1,ke2016online,ke2017efficient}, the main weakness is that a discarded point may be too far away from its corresponding line segment. So such a discarded point can not be approximately represented by its corresponding line segment well. On the other hand, PED, SED, LISSED and PSED are all position-based distances, which are defined based on the Euclidean distance between each discarded point and its "mapped" position on the compressed trajectory. But they do not provide any error guarantee on the direction information\cite{long2013direction}. For SED\cite{meratnia2004spatiotemporal,potamias2006sampling,muckell2014compression,muckell2011squish} and LISSED\cite{cao2017dots,6142066}, the time attribute of each discarded point is used to find its synchronized point on the corresponding line segment. Introduced in Section \ref{subsection:PEDandRPED}, PED\cite{douglas1973algorithms,keogh2001online,meratnia2004spatiotemporal,hershberger1992speeding,liu2015bounded,liu2016novel,lin2017one} is adopted by most existing line simplification methods\cite{lin2017one}, and there are mainly 4 most popular trajectory compression algorithms in online mode using PED as their error metric, i.e. OPW\cite{keogh2001online,meratnia2004spatiotemporal}, BQS\cite{liu2015bounded,liu2016novel}, FBQS\cite{liu2015bounded,liu2016novel} and OPERB\cite{lin2017one}. Among them, OPW is proposed the most early and it compresses a trajectory segment as long as possible into an $\epsilon$-error-bounded line segment, with $\epsilon$ being the upper bound of PED value. During each loop iteration to compress a trajectory segment into a line segment, BQS builds a virtual coordinate system centered at the starting point at the beginning. In each of four quadrants, BQS establishes a rectangular bounding box and two bounding lines so that in some cases, a point can be quickly decided for removal or preservation without needing expensive error calculation. For FBQS, a fast version of BQS, a raw trajectory point is directly reserved when error calculation is needed in BQS. So error calculation is no longer needed in FBQS. So, FBQS is a little faster than BQS at the expense of the larger size of the generated compressed trajectories. OPERB is based on a novel distance checking method and a directed line segment is used to approximate the buffered points. For BQS and FBQS, the accuracy loss of the compressed trajectories is relatively small, but their time costs are both extremely high. For OPERB and OPW, relatively less execution time is needed, but at the expense of the extremely high accuracy loss of the compressed trajectories. So ROCE, which makes the best balance among the accuracy loss, the time cost and the compression rate, is badly needed.

$\mathbf{Query\ on\ trajectories.}$ Large trajectory data facilitates various real-world applications, such as route planing, trajectory pattern mining and travel time prediction, and lots of attention has been drawn in queries on trajectories, such as \cite{duan2018bus,yuan2019distributed,shang2017trajectory,xu2019continuous,ali12maximum}. For example, by analyzing large amounts of historical trajectories, \citet{dai2016path} and \citet{dai2015personalized} study how to provide better navigation services for a driver on considering time cost, fuel consumption and the preference of the driver. In trajectories generated by the same person, there must be some common features hidden. And \citet{wu2016fuzzy} and \citet{jin2019moving} investigate the potential for historical trajectories accumulated from different sources to be linked so as to reconstruct a larger trajectory of a single person. Detecting anomalous trajectories (i.e. detours) has become an important and fundamental concern in many real-world applications. \citet{liu2020online} proposes a novel deep generative model to solve the problem of online anomalous trajectory detection. Real-time co-movement pattern mining for trajectories is to discover co-moving objects that satisfy specific spatio-temporal constraints in real time, and it serves a range of real-world applications, such as traffic monitoring and management. Targeting the visualization and interaction with such co-movement detection on streaming trajectories, \citet{fang2020coming} proposes a real-time co-movement pattern mining system to handle streaming trajectories. As more and more trajectories are being generated, the amount of trajectories to be queried usually exceeds the storage and processing capability of a single machine. But the situations, where the queried and analyzed trajectories are too much to be queried or they are all compressed trajectories, are considered in none of these works. To address these, \citet{shang2018dita} adopts a different strategy from us, and proposes a distributed in-memory trajectory analytics system to support large-scale trajectory analytics in distributed environments. This strategy and our queries on compressed trajectories are orthogonal to each other, and may be combined with each other to further improve the efficiency of queris on trajectory.

\section{Conclusions}\label{section:Conclusions}

For existing trajectory compression algorithms in online mode, either too much execution time is needed, or the accuracy loss of the compressed trajectories is not tolerable. To address this, this paper has presented ROCE, an efficient compression algorithm. A range query serves as a primitive, yet essential operation on trajectories. Using the traditional criteria on compressed trajectories will lead to that many trajectories will be missing in the result set. To solve this problem, we propose a new criteria based on the probability and an efficient range query processing algorithm RQC. An efficient index \emph{ASP\_tree} and lots of novel techniques are also presented to accelerate the processing of trajectory compression and range queries. Extensive experiments have been conducted using real-life trajectory datasets. The results demonstrate that ROCE makes the best balance among the accuracy loss, the time cost and the compression rate, and the difference between range query results on compressed trajectories and those on the corresponding raw trajectories is reduced greatly by using our RQC algorithm.

In the future, we consider to propose an optimal compression algorithm based on PSED, which can compress a trajectory into an $\epsilon$-error-bounded compressed trajectory with the smallest number of consecutive line segments. And this compression algorithm should also be of high-efficiency. On compressed trajectories, more kinds of queries should be also well studied to reduce the difference between the query results on compressed trajectories and those on the corresponding raw trajectories.

\bibliographystyle{spbasic} 
\bibliography{sample}

\end{document}